\documentclass[10pt,oneside,a4paper]{article}

\usepackage{microtype} 
\usepackage{soul}
\usepackage{upgreek}
\usepackage{enumitem}
\usepackage{subfigure}
\usepackage{booktabs}
\usepackage{multirow}
\usepackage{longtable}
\usepackage{microtype}

\usepackage{graphicx}
\usepackage{amsmath}
\usepackage{amssymb}
\usepackage[top=1in, bottom=1.25in, left=1.30in, right=1.30in]{geometry}
\usepackage{float}

\newcommand\figureHeight{56mm}
\newcommand\tableSpace{2mm}

\begin{document}
\pagenumbering{arabic}

\title{
Numerical Modelling of Satellite Downlink Signals in a Finslerian-Perturbed Schwarzschild Spacetime\\
\vspace{2mm}
\small{refereed version of this article: \textit{Universe} \textbf{2020}, \textit{6}, 57}}

\author{Ingo Abraham $^{1,\dagger}$, Wolfgang Hasse $^{1,2,*,\dagger}$ and Martin Plato $^{1,\dagger}$\vspace{2mm}\\
$^{1}$ Wilhelm Foerster Observatory Berlin,\\Munsterdamm~90, 12169~Berlin, Germany$\vspace{2mm}$\\
$^{2}$ Institute for Theoretical Physics, Technical University Berlin,\\Hardenbergstra{\ss}e~36, 10623~Berlin, Germany\vspace{2mm}\\
$^*$ astrometrie@gmx.de\\$^{\dagger}$ These authors contributed equally to this work.}

\date{April 17, 2020}

\maketitle


\begin{abstract}
The work presented in this paper aims to contribute to the problem of testing Finsler gravity theories by means of experiments and observations in the solar system. Within a class of spherically symmetric static Finsler spacetimes we consider a satellite with an on-board atomic clock, orbiting in the Finslerian-perturbed gravitational field of the earth, whose time signal is transmitted to a ground station, where its receive time and frequency are measured with respect to another atomic clock. This configuration is realized by the Galileo 5 and 6 satellites that have gone astray and are now on non-circular orbits. Our method consists in the numerical integration of the satellite's orbit, followed by an iterative procedure which provides the numerically integrated signals, i.e., null geodesics, from the satellite to the ground station. One of our main findings is that for orbits that are considerably more eccentric than the Galileo 5 and 6 satellite orbits, Finslerian effects can be separated from effects of perturbations of the Schwarzschild spacetime within the Lorentzian geometry. We also discuss the separation from effects of non-gravitational perturbations. This leads us to the conclusion that observations of this kind combined with appropriate numerical modelling can provide suitable tests of Finslerian modifications of general relativity.
\end{abstract}

\section{Introduction}\label{sec:intro}
{In this paper, we present a method of testing Finsler gravity, a generalization of Einstein's general theory of relativity, by means of experiments and observations in the solar system. We mainly focus on the case that the time signal of an atomic clock on board an Earth satellite is transmitted to a ground station, where it is compared with the time signal of another atomic clock. The Finslerian perturbations of the Schwarzschild spacetime affect the satellite orbit and the downlink signals and cause small shifts of the receive times and frequencies of the downlink signals, compared to the unperturbed Schwarzschild spacetime.}

Let us first make some remarks concerning the role of such experiments in testing the Einsteinian theory or generalized theories of gravity. One of the ``three classical tests of general relativity'' consists in verifying the gravitational redshift, which can be described as follows.
If a source of electromagnetic radiation is situated in a gravitational field, its spectrum measured outside or less deep in the field is shifted to longer wavelengths compared to its spectrum measured nearby the source. In other words, a standard clock in a gravitational field appears to run slower when observed from outside or less deep in the field. Assuming that atomic clocks are standard clocks, there is the possibility to test general relativity by comparing the time signals of two atomic clocks, one on board an Earth satellite and one on the ground.

Recently, the accuracy of such experiments has been considerably improved by using Galileo on-board atomic clocks. On 22 August 2014, the satellites Galileo 5 and 6 were unintentionally launched into eccentric orbits. From the point of view of gravitational physics, this mishap was a stroke of luck. Thanks to this, a little more than four years later, Delva~et~al.~\cite{reference:Delva_2018} and Hermann~et~al.~\cite{reference:Hermann_2018} were able to report on confirmation of the prediction from general relativity with an accuracy in the order of $10^{-5}$ at 1 $\sigma$ level. This is a few times more accurate than the Gravity Probe A experiment in 1976, see Vessot~et~al.~\cite{reference:Vessot_1980}, which had previously provided the most accurate confirmation of the general relativistic redshift effect. The accuracy now achieved opens up the possibility of not only testing Einstein's theory, but also of providing quantitative bounds on alternative theories of gravity by means of such experiments.

Finsler gravity is a generalization of general relativity, where the components of the metric tensor can depend not only on the spacetime coordinates but also on the components of a tangent vector. More precisely, the Lagrangian of the Finsler geodesics must no longer be bilinear in the components of the tangent vector but only fulfil the weaker requirement of homogeneity of degree two. Since bilinearity is a special case of homogeneity of degree two, Lorentzian geometry is a special case of Finsler geometry. Consequently, general relativity is a special case of Finsler gravity.

With regard to the position of Finsler gravity among the alternative theories of gravity, it should first be mentioned that it breaks spatial isotropy even on the tangent spaces. Therefore, the principle of local Lorentz invariance (LLI) is violated and the propagation of light in vacuum may be anisotropic.
Since LLI is part of the Einstein equivalence principle (EEP), the EEP is also violated in Finsler gravity. (As is well known, the validity of the EEP requires a pseudo-Riemannian spacetime geometry, see e.g., Will~\cite{reference:Will_2014}). However, since in our class of Finsler spacetimes (see Section~\ref{sec:Finsler_spacetime}) there is a unique timelike geodesic for every timelike initial condition, the weak equivalence principle is valid.

Since length is not a Lorentz invariant quantity, the above-mentioned violation of LLI can also be expected in a still to be found theory of quantum gravity which asserts that there is a fundamental length scale given by the Planck length. This motivates to consider non-quantized theories of Finsler gravity as possible ``interpolations'' between general relativity and quantum gravity in some regime of length and energy scales while maintaining the weak equivalence principle. More about this and other motivations, so in connection with the Ehlers-Pirani-Schild~\cite{reference:Ehlers_1972} axiomatic approach to general relativity, can be found in the recent review on the present status and the perspectives of Finslerian spacetime theories by L{\"a}mmerzahl and Perlick~\cite{reference:Laemmerzahl_2018} and in Section I of the article by L{\"a}mmerzahl, Perlick, and Hasse~\cite{reference:Laemmerzahl_2012}. The latter publication also provides the class of spherically {symmetric} static Finsler spacetimes on which our studies are based. The Finsler metrics of this model are purely kinematical perturbations of the Schwarzschild metric, i.e., these were not derived as solutions to any field equation.

Gravitational perturbations by higher multipole moments of the central body (a planet or the sun) or by other bodies (e.g., the sun and the moon in the case of a satellite orbiting the earth, the~planets in the case of a test particle in the gravitational field of the sun) as well as non-gravitational perturbations (e.g., solar radiation pressure) have not been taken into account. This is justified by the assumption that all these perturbations, as well as the Finslerian perturbations, are so small that in principle one may linearize the equations of motion with respect to the corresponding terms and treat all these perturbations separately, followed by addition of all their effects on measurable quantities. (In~this context, we note that in the already mentioned analyses~\cite{reference:Delva_2018, reference:Hermann_2018} of Galileo 5 and 6 data, to test the gravitational redshift predicted by Einstein's theory, from the here mentioned perturbations only Earth's oblateness, represented by the zonal coefficient $J_2$, has been considered.) Following this strategy{,} we restrict our analysis to perturbations which can be modelled by three ``perturbation functions'' which  {respect} the stationarity and the spherical symmetry of the spacetime. These perturbation functions will be introduced in the next section.

The paper is organized as follows.
In Section~\ref{sec:Finsler_spacetime}, we will fix our notations and conventions and introduce the considered class of Finsler spacetimes.
Section~\ref{sec:downlink_model} gives a brief description of our model of a downlink from an Earth satellite to a ground station.
The subsequent Section~\ref{sec:geodesic_equation} deals with the mathematical representation of our model and its numerical treatment.
Section~\ref{sec:LPH12} presents tests of our software against results published in~\cite{reference:Laemmerzahl_2012}.
In Section~\ref{sec:iterative_procedure} we explain the iterative procedure which provides the null geodesics of the downlink.
Readers who are less interested in the numerical aspects may skip Sections~\ref{sec:geodesic_equation}--\ref{sec:iterative_procedure}.
In order to separate the effects of the orbit perturbations from the effects of the signal perturbations we consider in Section~\ref{sec:hybrid_models} ``hybrid models'', in which both kinds of perturbations can be ``switched on'' and ``off'' separately.
The most important part of our paper is Section~\ref{sec:Galileo}. It~contains core findings obtained by means of our simulations. We discuss, exemplarily for the orbits of the Galileo~5 and 6 satellites, to which extent Finslerian effects can be separated from effects of perturbations of the Schwarzschild spacetime within the Lorentzian geometry.
Based on these results, we consider, in Section~\ref{sec:high_eccentricity_model}, a model satellite with high orbit eccentricity.
In the concluding Section~\ref{sec:conclusions}, we discuss, among other things, the advantages of combining the two Galileo satellites with a future dedicated mission and provide an outlook how our numerical approach can be brought over to other  {situations of communication} or even to other modifications of general relativity.

\section{A Class of Spherically Symmetric Static Finsler Spacetimes}\label{sec:Finsler_spacetime}

We work in the same class of Finsler spacetimes as in the paper of L{\"a}mmerzahl, Perlick, and Hasse~\cite{reference:Laemmerzahl_2012}. This class is defined in Schwarzschild coordinates $t$, $r$, $\vartheta$, and $\varphi$ by the Finsler Lagrangian
\begin{equation}\label{Lagrangian_A}
2\mathcal{L}=
(1+\phi_0)h_{tt}\dot{t}^2
+(1+\phi_1)h_{rr}\dot{r}^2
+r^2(\dot\vartheta^2+(\sin^2\vartheta)\dot{\varphi}^2)
+\phi_2\frac{h_{rr}r^2\dot{r}^2(\dot\vartheta^2 +(\sin^2\vartheta)\dot{\varphi}^2)}
{h_{rr}\dot{r}^2+r^2(\dot\vartheta^2 +(\sin^2\vartheta)\dot{\varphi}^2)},
\end{equation}
see (14) in~\cite{reference:Laemmerzahl_2012} or (37) in~\cite{reference:Hasse_2019},
which defines a perturbed Schwarzschild spacetime.
Here $h_{tt}$ and $h_{rr}$ are the Schwarzschild metric coefficients, given by
\begin{equation}
h_{tt} = -\dfrac{c^2}{h_{rr}} = -c^2\biggl(1 - \dfrac{2GM}{c^2r}\biggr) {= -c^2\biggl(1 - \dfrac{r_S}{r}\biggr)},
\end{equation}
where $c$ is the speed of light, $G$ is the gravitational constant,  $M$ is the mass of the gravitating body{, and $r_S=2GM/c^2$ is its Schwarzschild radius}.  The metric convention is ({-} + + +), i.e., the unperturbed metric has signature +2.

$\phi_0$, $\phi_1$, and $\phi_2$ are ``perturbation functions'' which depend differentiabl{y} only on the radial coordinate $r$.
For stationary observers, $\phi_0$ and $\phi_1$ perturb the measurement of proper time and radial length, respectively.
Whereas $\phi _0$ and $\phi_1$ describe perturbations within the Lorentzian geometry, $\phi_2$~introduces a spatial anisotropy which is a genuine Finsler feature.

Let us remark that it results from sections III and IV of~\cite{reference:Laemmerzahl_2012} that the choice of the Finsler Lagrangian \eqref{Lagrangian_A} is in some sense natural. To be more precise, it results from adding the leading-order spherical-harmonics term in a general Finslerian perturbation of the spatial part of the metric to the Lagrangian of the Schwarzschild metric, followed by linearisation with respect to the perturbing terms. Thereby, ``general'' means only restricted to compatibility with the spherical symmetry.

Throughout this paper, as in~\cite{reference:Laemmerzahl_2012}, we assume that the perturbation functions are so small that we may linearize all equations with respect to the $\phi_n(r)$ and their derivatives $\phi^\prime_n(r)$, $n$ = 0, 1, 2. The~outputs of our numerical codes are also based on this linearisation procedure.

In~\cite{reference:Laemmerzahl_2012} only problems are considered that can be reduced to the equatorial plane $\vartheta = \pi/2$ by spherical symmetry.
But in our case, when it comes to modelling the downlink, all three spatial dimensions must be taken into account.

The first order equations of motion in the plane $\vartheta=\pi/2$, with the constants of motion as parameters, are derived in Section IV of reference~\cite{reference:Laemmerzahl_2012}. From that one can easily derive the equations of motion for given initial values in the four-dimensional spacetime by means of a suitable spatial rotation of the coordinate system.

For later reference we note how $\phi_0$, $\phi_1$, and $\phi_2$ affect the propagation of material bodies and light rays as seen from a stationary observer at infinity (at least in linear approximation). This can be seen from the equations of motions in the Appendixes~\ref{app:2d_equations} and~\ref{app:3d_equations}. In general, $\phi_0$, $\phi_1$, and $\phi_2$ affect both the radial and the tangential velocity components. However, there are some special cases:

\begin{itemize}[leftmargin=*,labelsep=5.8mm]
\item	$\phi_0$, $\phi_1$, and $\phi_2$ never cause tangential acceleration components if the velocity is pure tangential or pure radial.
\item	$\phi_1$ always causes pure radial acceleration.
\item	$\phi_2$ does not affect pure radial velocities.
\end{itemize}

\section{Our Model of a Downlink from an Earth Satellite to a Ground Station}\label{sec:downlink_model}

Let us assume that the geometry outside the earth is determined by the Finsler Lagrangian \eqref{Lagrangian_A} where, in this case, $M$ is the earth's mass. We consider an Earth satellite whose orbit is modelled by a spatially bounded timelike geodesic. An on-board clock is modelled by the satellite's Finsler proper time (with arbitrary zero point).
Furthermore, we consider a ground station on the terrestrial surface. The rotation of the earth around its own axis is taken into account by assuming that the worldline of the ground station is a spatially circular timelike curve with constant value of the angular coordinate $\vartheta$, given by the station's latitude $\pi/2 - \vartheta$. The effect of the earth's rotation on the spacetime geometry is not taken into account. The model also contains a clock at the ground station which gives the station's Finsler proper time. Finally, the downlink from the satellite to the ground station is described by null geodesics connecting the world lines of the satellite and the ground station.

\section{Equations and Numerical Treatment of Our Model}\label{sec:geodesic_equation}

The equations of the causal geodesics, proper times, and frequency shifts that are the basis of the simulations of the satellite orbits and the signal propagation presented in Sections~\ref{sec:Galileo} and~\ref{sec:high_eccentricity_model} are given in Appendix~\ref{app:3d_equations}.
The iterative procedure for determining the null geodesics from the satellites to the ground stations will be outlined in Section~\ref{sec:iterative_procedure}.

The calculations were performed by one of us (I.A.) with the {software Mathematica}\textsuperscript{\textregistered}, version 12.0.0.0, making use of its inbuilt procedures for numerical (differential) equation solving and integration. The numerical effort was reduced by taking advantage of the fact that every causal geodesic $\mathbf{r}(t)$ is restricted to the plane spanned by the vectors $\mathbf{r}(t_0)$ and $d\mathbf{r}/dt(t_0)$ at arbitrary time $t_0$. Therefore, all satellite orbits and signals (null geodesics) were calculated by choosing two orthonormal basis vectors $\mathbf{k}$ and $\mathbf{l}$ in the plane spanned by the respective initial position and velocity vectors and then solving the two-dimensional analogue of \eqref{acceleration_vector_a}. The working precision (maximum number of digits for internal computations) was 105 for the satellite orbits and 65 for all other calculations.
%
\section{Comparison with the Results of L{\"a}mmerzahl, Perlick, and Hasse}
{In} the article by L{\"a}mmerzahl, Perlick, and Hasse~\cite{reference:Laemmerzahl_2012} the effects of $\phi_0$, $\phi_1$, and $\phi_2$ on three known phenomena (time delay of light rays, light~deflection by the sun, perihelion precession of the planet Mercury) are presented {under the assumption that with regard to these effects the perturbation functions can be approximated as (see~(39) in~\cite{reference:Laemmerzahl_2012}\label{sec:LPH12})}

\begin{equation}\label{phi_n1_a}
{\phi_{n} = \phi_{n1}\dfrac{r_S}{r},\ n=0,\ 1,\ 2}.
\end{equation}

These results were obtained by numerical treatment of integrals rather than by numerical treatment of the geodesic equations. It is therefore an obvious approach, among other tests, to~calculate these effects also with the software described in the preceding section and to compare the results.

In general the calculations described in Section~\ref{sec:geodesic_equation} confirm the results presented in~\cite{reference:Laemmerzahl_2012}, but we found two discrepancies regarding light deflection and perihelion precession. Therefore, we additionally calculated these two effects on basis of the equations in Appendix~\ref{app:2d_equations}. These calculations were performed again by one of us (I.A.) with Mathematica and independently by another of us (M.P.) with the {software MATLAB}\textsuperscript{\textregistered}, version 5.3.0.10183 (R11). The MATLAB calculation made use of its available solver functions for ordinary differential equations and were performed with the relative tolerance of the solver output set to $10^{-12}$.
%
\subsection{Time Delay of Light Rays}

Consider a radio signal that is sent from Earth ({r} = 1 au){, then} passes the sun at r = 0.0074 au and finally reaches a spacecraft at r = 8.43 au. According to Equation (82) in~\cite{reference:Laemmerzahl_2012} the difference between the perturbed one-way time delay $\delta t$ and the unperturbed one-way time delay $\delta t_0$ is
\begin{equation}
\delta t - \delta t_0 = (6.6\times 10^{-5}\ \phi_{11} + 3.3\times 10^{-6}\ \phi_{21} - 7.5\times 10^{-5}\ \phi_{01})s.
\end{equation}

The calculations described in Section~\ref{sec:geodesic_equation} exactly confirm this result.

\subsection{Light Deflection}\label{sec:LPH12_light_deflection}

Consider a light ray from a distant star (r $\rightarrow\infty$) that grazes the sun at r = 0.0046 au and finally reaches Earth (r = 1 au). According to Equation (73) in~\cite{reference:Laemmerzahl_2012} the difference between the perturbed deflection angle $\Delta\varphi$ and the unperturbed deflection angle $\Delta\varphi _0$ is

\begin{equation}\label{test_2}
\Delta\varphi - \Delta\varphi _0 = 4.2\times 10^{-6}\ \phi_{11} + 5.1\times 10^{-11}\ \phi_{21} - 4.2\times 10^{-6}\ \phi_{01}.
\end{equation}

\textls[-15]{Regarding the coefficients of $\phi_{01}$ and $\phi_{11}$ our calculations confirm Equation \eqref{test_2}, with maximum deviation $\pm$ 1 in the last reported digit. But for the coefficient of $\phi_{21}$ the results agree less well.
For~this coefficient, both Mathematica programs give the value $3.6\times 10^{-11}$ (for e.g., $10^{-8} \leq \phi_{21} \leq10^{-2}$)}. An~exemplary calculation with MATLAB gives the value $2.9\times 10^{-11}$. We cannot explain this discrepancy at the moment.

\subsection{Perihelion Precession}\label{sec:LPH12_perihelion_precession}

Equation (107) in~\cite{reference:Laemmerzahl_2012} for the perihelion motion of Mercury is
 
\begin{equation}\label{Mercury}
\dfrac{\omega-\omega _0}{\omega _0} = 3.3\times 10^{-1}\ \phi_{11} + 3.1\times 10^{-1}\ \phi_{21} + 2.5\times 10^{-8}\ \phi_{01}.
\end{equation}

In this equation $\omega$ and $\omega_0$ denote the precession rate (``angular velocity'') of the perihelion in the perturbed and unperturbed case, respectively. Whereas our calculations confirm the coefficients of $\phi_{11}$ and $\phi_{21}$, with maximum deviation $\pm$ 1 in the last reported digit, for the coefficient of $\phi_{01}$ we get the value of $-$0.5, which differs by more than seven powers of ten from the value in Equation \eqref{Mercury}.
We cannot uncover the cause of this discrepancy, but can only guess that it has to do with the orbital period. Namely, since the perihelion motion is calculated with fixed apsidal distances, $\phi_0$, $\phi_1$, and $\phi_2$ cause a change ${\Delta}T_s$ of the sidereal period, whereby the effect of $\phi_0$ clearly dominates (on circular orbits the effects of $\phi_1$ and $\phi_2$ even vanish, cf. Equation (30) in~\cite{reference:Laemmerzahl_2012}). ${\Delta}T_s$ in turn makes the largest contribution to the change ${\Delta}T_a$ of the anomalistic period. Ultimately, ${\Delta}T_a$ makes the dominating contribution to $\Delta\omega$. This contribution can be calculated analytically in the framework of the PPN formalism, which also results in a value of approximately $-$0.5 for the coefficient of $\phi_{01}$.

\section{The Iterative Procedure for Determining the Downlink Signals}\label{sec:iterative_procedure}

In our model the satellite transmits a signal to the ground station every 2000 s with respect to its proper time $\tau$. Therefore, after the calculation of the satellite orbit $\mathbf{r}_S(t)$, the next step is the calculation of the coordinate transmit times $t_n$ corresponding to the proper transmit times $\tau_n$ by integration of~\eqref{proper_time}. Then the null geodesics from the satellite positions $\mathbf{r}_S(t_n)$ to the rotating ground station with position $\mathbf{r}_B(t)$ are calculated by means of \eqref{acceleration_vector_a}. To this aim the main task is the iterative determination of the unknown emission directions $\tilde{\mathbf{d}}(t_n)$. We will explain the procedure for a signal emitted at $t_0$ with unknown emission direction $\tilde{\mathbf{d}}(t_0)$.

The first approximation $\mathbf{d}_1$ of the emission direction is calculated according to the formula
\begin{equation}
\mathbf{d}_1 := \mathbf{r}_B\biggl(t_0 + \frac{\lvert\mathbf{r}_B(t_0) - \mathbf{r}_S(t_0)\rvert}{c}\biggr) - \mathbf{r}_S(t_0).
\end{equation}

By making use of \eqref{lightspeed} for the coordinate light speed $\overline{c}$ at the emission position in the emission direction, the signal $\mathbf{r}_L(t)$ emitted at the position $\mathbf{r}_S(t_0)$ in the direction $\mathbf{d}_1$ is then calculated, together with its approach distance $\Delta:=\lvert\mathbf{r}_L(t)-\mathbf{r}_B(t)\rvert_{min}$ from the ground station.
Next the gradient of the approach distance with respect to the emission direction is calculated, i.e.,
\begin{equation}
\frac{\partial\Delta}{\partial\mathbf{d}}(\mathbf{d}_1),
\end{equation}
and the emission direction is updated according to
\begin{equation}
\mathbf{d}(s) := \mathbf{d}_1 - s\ \frac{\partial\Delta}{\partial\mathbf{d}}(\mathbf{d}_1)
\end{equation}
for a series of steps of increasing width $s$, until the approach distance stops decreasing at a step width $s_{max}$. This defines the improved emission direction
\begin{equation}
\mathbf{d}_2 := \mathbf{d}(s_{max}).
\end{equation}

This strategy is repeated several times and after the calculation of typically about 40 null geodesics the approach distance is typically only about 5 $\upmu$m.
The time $t_a$ of closest approach is used as coordinate receive time $t_r$ of the signal at the ground station. By integration of \eqref{proper_time} the coordinate receive times are converted to the proper receive times $\tau_r$ at the ground station.
Finally, for every signal the receive frequency shift is calculated according to \eqref{frequency_shift_2}.

\section{Hybrid Models}\label{sec:hybrid_models}

Since the perturbations influence both the orbital motion of the satellite and the signal propagation, the interesting question arises which of the two influences predominates in typical cases. This is not at all clear from the outset, as the following considerations show.

The orbital speeds are about five powers of ten smaller than the speed of light, so the satellites ``feel'' the $\phi_1$- and the $\phi_2$-perturbations (affecting only the spatial geometry) much less than the signals. One would therefore expect that the observable effects of the $\phi_1$- and the $\phi_2$-perturbations are quite predominantly caused by perturbations of the signal propagation, while a $\phi_0$-perturbation should affect the satellite orbit and the signals approximately equally.
However, this expectation does not take into account that the perturbations act on the signals only for a short time and that the null geodesics describing the signals and the timelike geodesics describing the satellite orbits must fulfil significantly different initial and boundary conditions. While for the satellite orbit the initial position and the initial velocity are given, for the signals the initial position and the condition to reach the ground station are given. Thus, in a certain sense the geodesic of the signal is ``more closely tied'' than that of the satellite and therefore less ``bent'' by the perturbations. In addition, the effects of the perturbations on the signals are never accumulative. The situation is different with orbital perturbations because even if a disturbed orbit is compared with an undisturbed orbit with the same sidereal period, there is an accumulative effect due to the generally different rates of perihelion precession.

Further consideration shows that the so far open question of the predominant effect is relevant with regard to the separation of Finslerian effects from effects of other perturbations. In our class of models, Finslerian ($\phi_2$-) perturbations do not affect satellites and signals in radial motion and their effects are greatest in a certain angular range between radial and tangential motion. Therefore, if~the orbit perturbations dominate, Finslerian effects would be greatest in certain sections of the orbit between the apsides and would increase with the eccentricity of the orbit. Conversely, if the signal perturbations dominate, Finslerian effects would be greatest when the satellite, seen from the ground station, is in a certain middle range of altitude above the horizon.

In order to answer this question we additionally studied ``hybrid models'', in which only the satellite orbit is and is not, respectively, affected by the perturbations. These models are not intended as a description of nature. Rather, they help to understand the results obtained with the actual model and to identify those orbits, orbit segments, and constellations (relative positions) of the satellite and the ground station that are most promising for the detection of possible genuine Finslerian effects.

\section{Galileo 5 and 6}\label{sec:Galileo}

\subsection{Simulation Parameters}
We will now study the effects of the first order perturbations $\phi_n(\rho)=\phi_{n1}\ r_S/\rho\ ({n} = 0,\ 1,\ 2)$ on the time signals that are transmitted from the satellites Galileo 5 and 6 and received at a ground station on Earth.
To that aim we simulated the motion of two model satellites with the following orbital parameters, that are similar to the parameters of the satellites Galileo 5 and 6: inclination $i=50^\circ$, perigee distance \mbox{$\rho_p$ = 23,500 km}, and perigee speed $v_p=4.436$ km/s and thus apogee distance \mbox{$\rho_a$ = 32,462 km}, semi-major axis $a$ = 27,981 km, and eccentricity $e=0.1601$.
As the arguments of perigee $\omega$ of the satellites Galileo 5 and 6 (GSAT0201 and GSAT0202) change with a rate of $0.034^\circ/d$~\cite{reference:gsc} the simulations were performed both for $\omega=90^\circ$ and $\omega=180^\circ$.

The signals from the satellites were received at two ground stations that followed the rotation of the earth.
For the description of the (spherical) earth we used the parameter values radius \mbox{$\rho_E = 6371$ km}, angular speed $\omega_E = d\varphi/dt = 7.292115\times 10^{-5}$ rad/s, standard gravitational parameter $GM_E$ = 398,600.44 km$^3$/s$^2$, and Schwarzschild radius $r_S = 2 GM_E/c^2$.
The first ground station was on the equator and received the signals from the satellite with argument of perigee $\omega=90^\circ$. A second ground station was at a latitude of $45^\circ$ north and received the signals from the satellite with argument of perigee $\omega=180^\circ$. The detailed simulation parameters are listed in Table~\ref{table:galileo_model_parameters}.

As we will find out, each perturbation causes characteristic patterns in the received signals, independent of the argument of perigee of the satellite and the latitude of the ground station. 

To check the reliability of the results, the simulations were performed with two different values of each parameter $\phi_{n1}$. We will see, that all results {linearly} depend on the parameters $\phi_{n1}$ in the time range under study, to a very good approximation. This confirms both the assumed smallness of the perturbations and the reliability of the results.

\begin{table}[H]
\caption{Simulation parameters: argument of perigee $\omega$ and orbital state vectors $\mathbf{r}$ and $\mathbf{v}$ of satellite S and ground station B at initial time $t_0$. $\mathbf{\hat{x}}$, $\mathbf{\hat{y}}$, and $\mathbf{\hat{z}}$ denote the unit vectors along the coordinate axes.\vspace{\tableSpace}}
\label{table:galileo_model_parameters}
\centering

\setlength{\tabcolsep}{2.0mm}
\begin{tabular}{c r@{\hskip 0.8mm}l r@{\hskip 0.8mm}l r@{\hskip 0.8mm}l r@{\hskip 0.8mm}l c}
\toprule
$\omega$&
\multicolumn{2}{c}{$\mathbf{r}_B(t_0)$}&
\multicolumn{2}{c}{$\mathbf{v}_B(t_0)$}&
\multicolumn{2}{c}{$\mathbf{r}_S(t_0)$}&
\multicolumn{2}{c}{$\mathbf{v}_S(t_0)$}&
\multicolumn{1}{c}{$t_0$\ [s]}\\
\midrule
$90^\circ$&
$\rho_E$&$\mathbf{\hat{x}}$&
$\rho_E\omega_E$&$\mathbf{\hat{y}}$&
$\rho_p$&$(\cos i\ \mathbf{\hat{x}}+\sin i\ \mathbf{\hat{z}})$&
$v_p$&$\mathbf{\hat{y}}$&
$0$\\[1.0ex]
$180^\circ$&
$\dfrac{\rho_E}{\sqrt{2}}$&$(\mathbf{\hat{x}}+\mathbf{\hat{z}})$&
$\dfrac{\rho_E}{\sqrt{2}}\omega_E$&$\mathbf{\hat{y}}$&
$\rho_p$&$\mathbf{\hat{y}}$&
$-v_p$&$(\cos i\ \mathbf{\hat{x}}+\sin i\ \mathbf{\hat{z}})$&
$250,000$\\
\bottomrule
\end{tabular}
\end{table}

\subsection{Sidereal Period}\label{subsec:sidereal period}

In the following sections we will compare the received signals at the ground stations in the unperturbed spacetime and in the perturbed spacetimes. For the underlying satellite orbits we require the same value of the ``area coordinate'' $r$ at perigee (same value of $\rho_p$) and the same sidereal period, measured with respect to the proper time of a stationary clock far away from the earth. The proper time of such a clock can be sufficiently approximated by the coordinate time $t$. With respect to the coordinate time the sidereal period in the unperturbed spacetime is $T_0$ = 46,580.92981316189761~s. Keeping the orbital state vectors of the satellites at $t_0$ fixed according to Table~\ref{table:galileo_model_parameters}, the perturbation functions $\phi_n$ result in shifts $\Delta T=T-T_0$ of the sidereal period that are listed in Table~\ref{table:galileo_sidereal_periods}.

\begin{table}[H]
\caption{Shifts $\Delta T$ of the sidereal periods of the model satellites, divided by $T_0$, with orbital state vectors at $t_0$ fixed according to Table \ref{table:galileo_model_parameters}.\vspace{\tableSpace}}
\label{table:galileo_sidereal_periods}
\centering
\setlength{\tabcolsep}{2.0mm}
\begin{tabular}{l l r@{\hskip 0.8mm}c@{\hskip 0.8mm}l}
\toprule
$n$	&	\multicolumn{1}{c}{$\phi_{n1}$}	&	\multicolumn{3}{c}{$\dfrac{\Delta T}{T_0}$}\\
\midrule
0		&	$10^{-9}$												&	$2.57$	& $\times$	& $10^{-9}$\\
0		&	$10^{-8}$												&	$2.57$	& $\times$	& $10^{-8}$\\
1		&	$0.1$														&	$4.22$	& $\times$	& $10^{-12}$\\
1		&	$1$															&	$4.22$	& $\times$	& $10^{-11}$\\
2		&	$0.1$														&	$4.06$	& $\times$	& $10^{-12}$\\
2		&	$1$															&	$4.06$	& $\times$	& $10^{-11}$\\
\bottomrule
\end{tabular}
\end{table}

Table~\ref{table:galileo_sidereal_periods} shows, that in the parameter range under study and with orbital state vectors at $t_0$ fixed according to Table~\ref{table:galileo_model_parameters}, the shift of the sidereal period {linearly} depends on the parameters $\phi_{n1}$ according~to
\begin{equation}
\dfrac{\Delta T}{T_0}=2.6\ \phi_{01}+4.2\times 10^{-11}\ \phi_{11}+4.1\times 10^{-11}\ \phi_{21}.
\end{equation}

Therefore, to fulfil the constraint of same sidereal periods in the unperturbed spacetime and in the perturbed spacetimes, the simulations were performed with a slightly reduced satellite speed at $t_0$ in the perturbed spacetimes according to Table~\ref{table:galileo_velocity_correction}.

\begin{table}[H]
\caption{Residual shifts $\Delta T$ of the sidereal periods of the model satellites, divided by $T_0$, with reduced satellite speeds $v_p + \Delta v_p$ at $t_0$.\vspace{\tableSpace}}
\label{table:galileo_velocity_correction}
\centering
\setlength{\tabcolsep}{2.0mm}
\begin{tabular}{l l r@{\hskip 0.8mm}c@{\hskip 0.8mm}l r@{\hskip 0.8mm}c@{\hskip 0.8mm}l}
\toprule
$n$	&	\multicolumn{1}{c}{$\phi_{n1}$}	&	\multicolumn{3}{c}{$\dfrac{\Delta v_p}{v_p}$}	&	\multicolumn{3}{c}{$\dfrac{\Delta T}{T_0}$}\\
\midrule
0		&	$10^{-9}$												&	$-6.20653150$	& $\times$	& $10^{-10}$				&	$1.49$	& $\times$	& $10^{-18}$\\
0		&	$10^{-8}$												&	$-6.20653152$	& $\times$	& $10^{-9}$					&	$-1.11$	& $\times$	& $10^{-17}$\\
1		&	$0.1$														&	$-1.01973$		& $\times$	& $10^{-12}$				&	$2.53$	& $\times$	& $10^{-18}$\\
1		&	$1$															&	$-1.01973$		& $\times$	& $10^{-11}$				&	$2.53$	& $\times$	& $10^{-17}$\\
2		&	$0.1$														&	$-9.800645$		& $\times$	& $10^{-13}$				&	$-4.19$	& $\times$	& $10^{-20}$\\
2		& $1$															&	$-9.800645$		& $\times$	& $10^{-12}$				&	$-4.15$	& $\times$	& $10^{-19}$\\
\bottomrule
\end{tabular}
\end{table}

\subsection{Satellite Orbits}
\begin{figure}[H]
\centering
\includegraphics[height=\figureHeight]{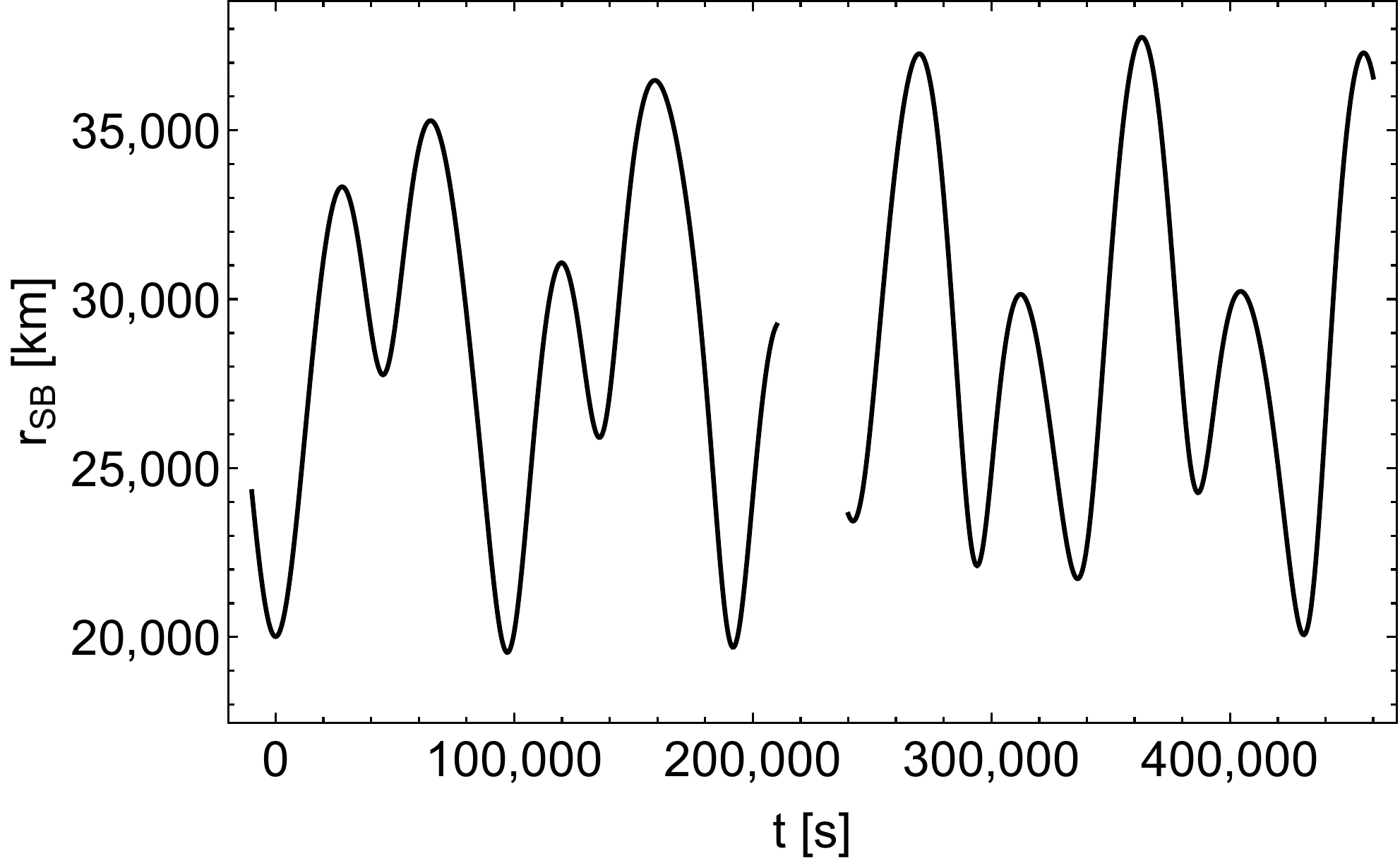}
\caption{{Coordinate} distance $r_{SB}:=\lvert\mathbf{r}_S-\mathbf{r}_B\rvert$ between the satellite and the ground station in the Schwarzschild spacetime, left (right) curve $\omega$ = $90^\circ$ ($180^\circ$).}
\label{figure:galileo_distance_satellite_base_1}
\end{figure}

In this section we report about the effects of the perturbation functions on the satellite orbits. To~start with, Figure~\ref{figure:galileo_distance_satellite_base_1} shows the coordinate distance between the satellite and the ground station in the unperturbed spacetime as function of the coordinate time.

\begin{figure}[H]
\centering
\includegraphics[height=\figureHeight]{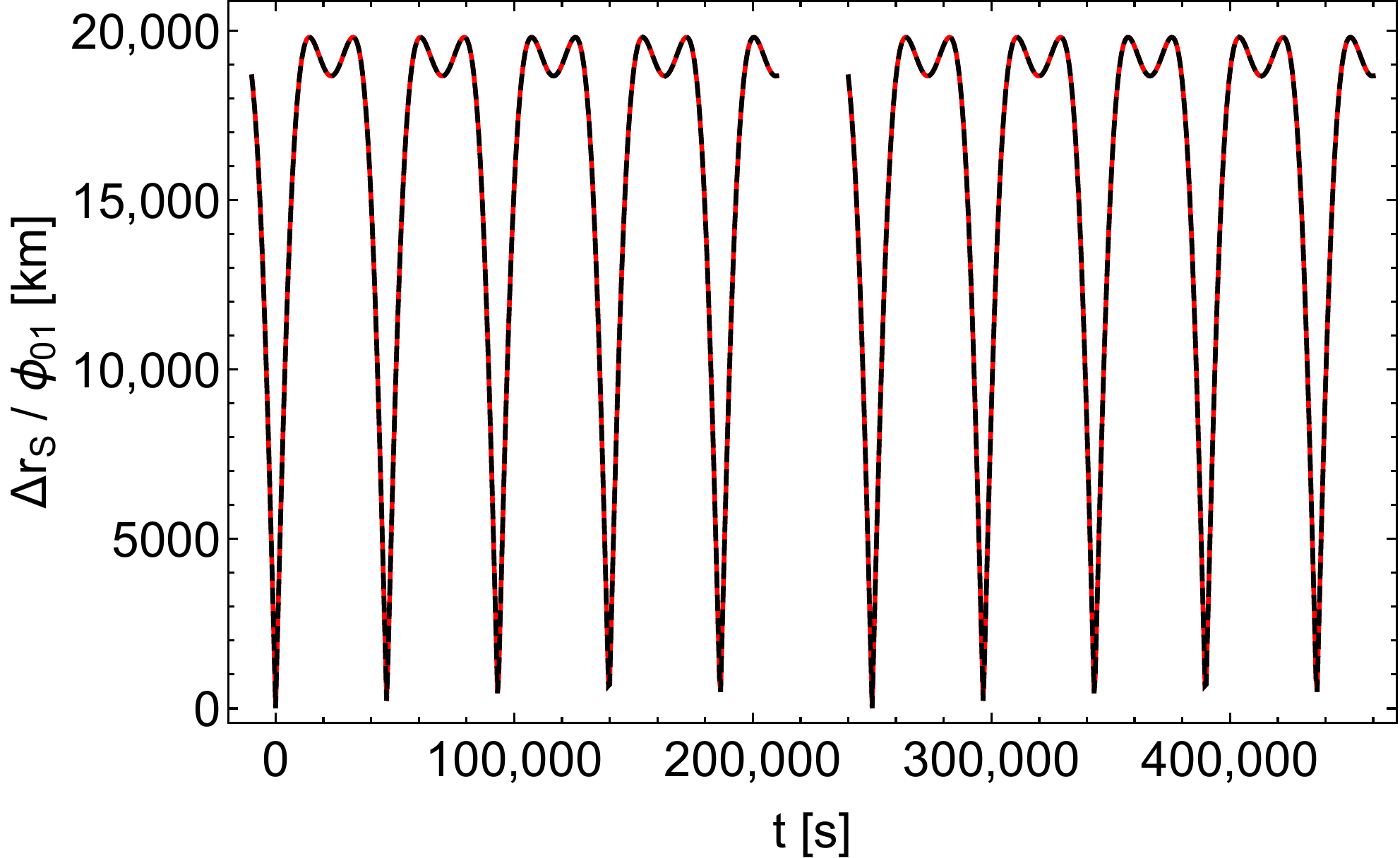}
\caption{Shift $\Delta r_S:=\lvert\mathbf{r}_S(\phi_{01}\neq 0)-\mathbf{r}_S(\phi_0=0)\rvert$ of the satellite position $\mathbf{r}_S$, divided by $\phi_{01}$. Black~dashes $\phi_{01}=10^{-9}$, red dashes $\phi_{01}=10^{-8}$, left (right) curves $\omega$ = $90^\circ$ ($180^\circ$).}
\label{figure:galileo_shift_satellite_position_0}
\end{figure}

\begin{figure}[H]
\centering
\includegraphics[height=\figureHeight]{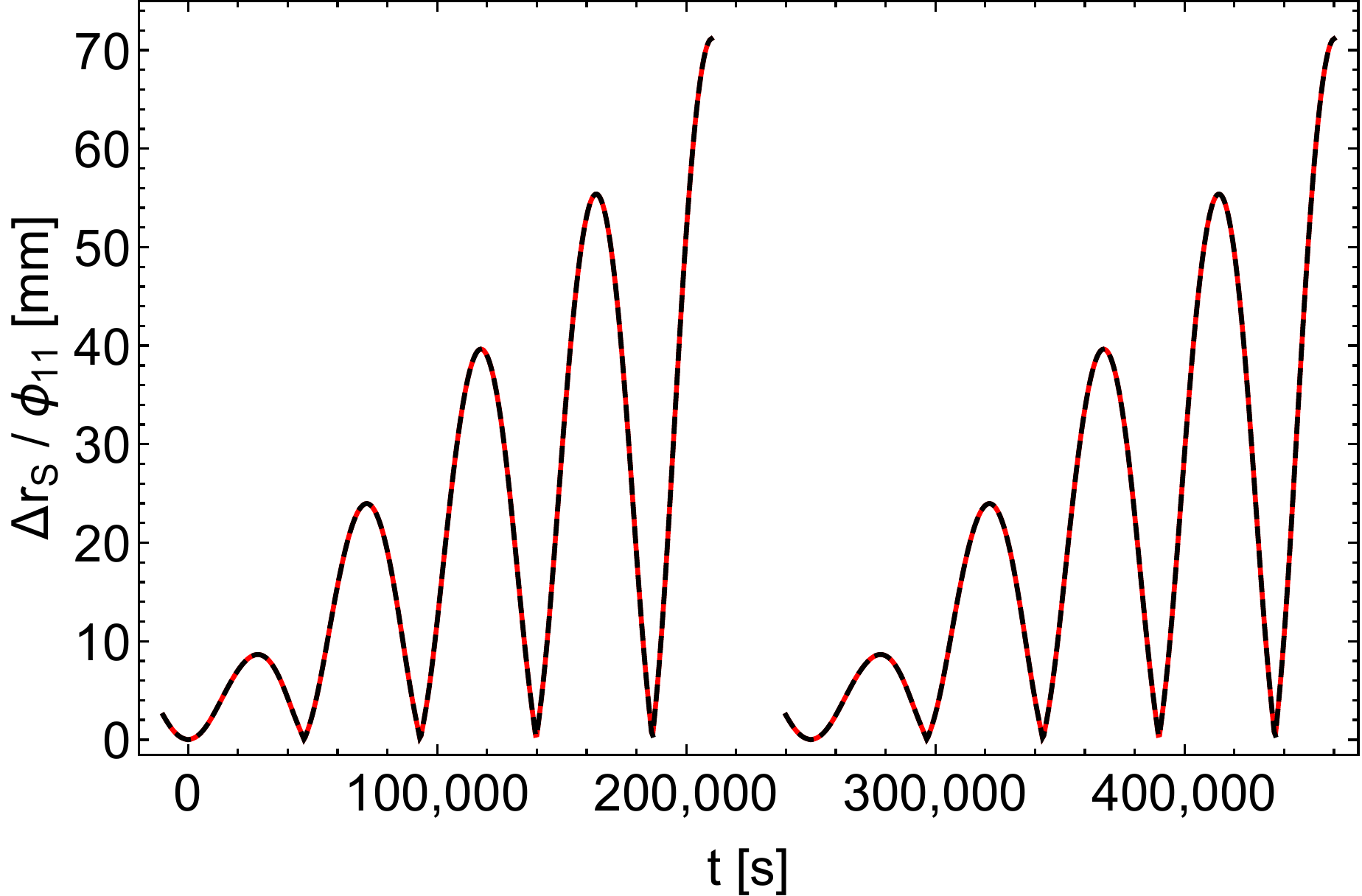}
\caption{Shift $\Delta r_S:=\lvert\mathbf{r}_S(\phi_{11}\neq 0)-\mathbf{r}_S(\phi_1=0)\rvert$ of the satellite position $\mathbf{r}_S$, divided by $\phi_{11}$. Black~dashes $\phi_{11}=0.1$, red dashes $\phi_{11}=1$, left (right) curves $\omega$ = $90^\circ$ ($180^\circ$).}
\label{figure:galileo_shift_satellite_position_1}
\end{figure}

\begin{figure}[H]
\centering
\includegraphics[height=\figureHeight]{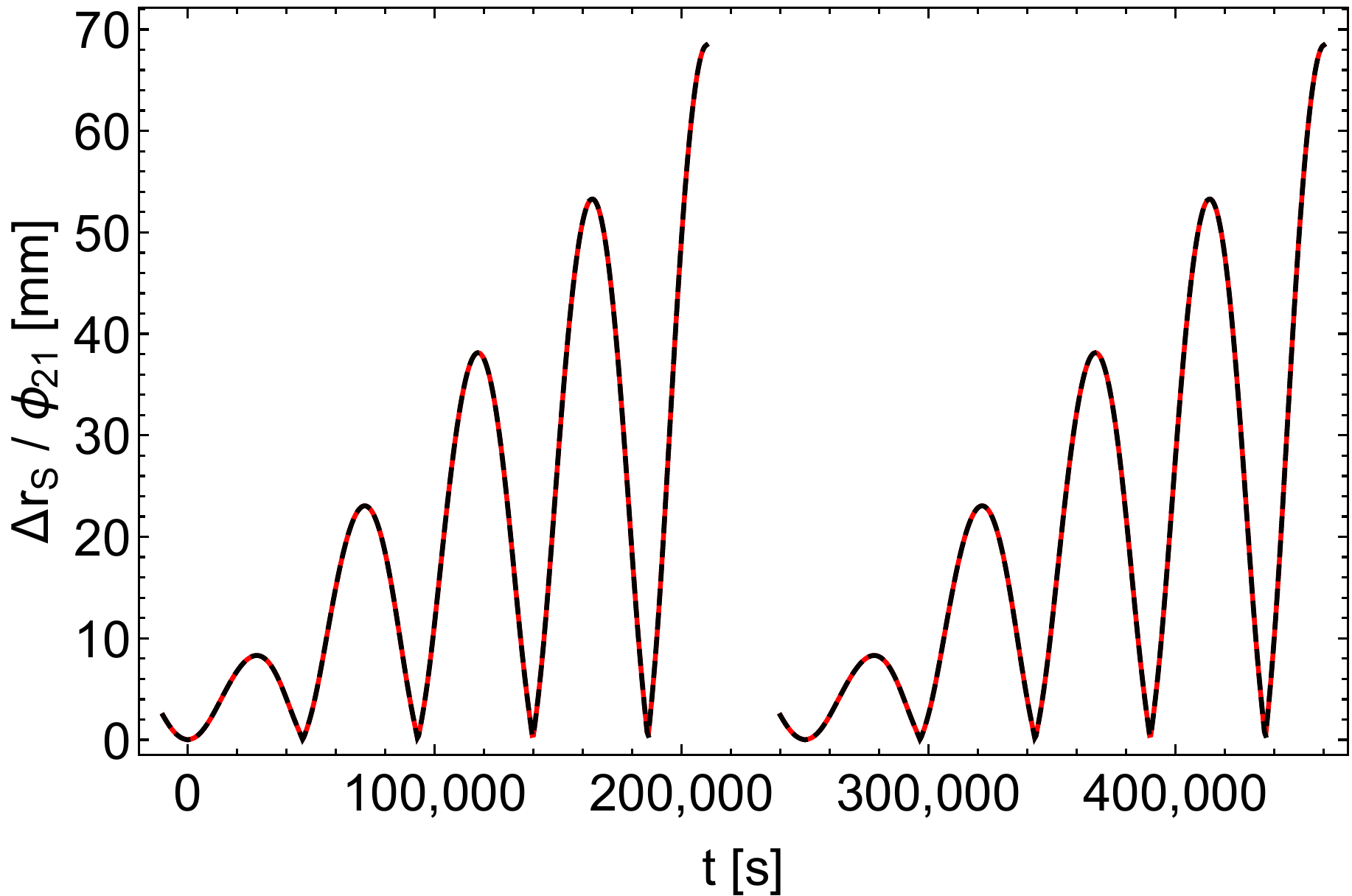}
\caption{Shift $\Delta r_S:=\lvert\mathbf{r}_S(\phi_{21}\neq 0)-\mathbf{r}_S(\phi_2=0)\rvert$ of the satellite position $\mathbf{r}_S$, divided by $\phi_{21}$. Black~dashes $\phi_{21}=0.1$, red dashes $\phi_{21}=1$, left (right) curves $\omega$ = $90^\circ$ ($180^\circ$).}
\label{figure:galileo_shift_satellite_position_2}
\end{figure}

In comparison, Figures~\ref{figure:galileo_shift_satellite_position_0}--\ref{figure:galileo_shift_satellite_position_2} show the shifts of the satellite positions due to the perturbation functions $\phi_n$, divided by the magnitudes $\phi_{n1}$ of the perturbations. In combination with the values of the parameters $\phi_{n1}$ listed in Table~\ref{table:galileo_velocity_correction} it can be seen from these figures, that in the depicted time span the maximum shifts of the satellite positions caused by the perturbations are 20 cm ($\phi_{01} = 10^{-8}$) and 7~cm ($\phi_{11}=\phi_{21} = 1$), respectively.

\unskip
\unskip
\unskip

$\phi_0$, $\phi_1$, and $\phi_2$ shift the satellite positions qualitatively different. $\phi_0$ causes an almost periodic pattern with almost constant amplitude.
In contrast, $\phi_1$ and $\phi_2$ cause patterns with increasing amplitudes, that are very similar to each other (the maximum shift of the satellite position is 72 and 69 mm, respectively). From the satellite orbits, subject to perturbations $\phi_{11}=1$ and $\phi_{21}=1$, respectively, it can be calculated that the distance between the two satellites increases in time, similar to Figures~\ref{figure:galileo_shift_satellite_position_1} and~\ref{figure:galileo_shift_satellite_position_2}, up to a maximum distance of 2,8 mm at the end of the depicted time span.

The root of the very similar effects of the two perturbations can be read from \eqref{acceleration_vector_a} and \eqref{f_of_t}. According to these equations, the differences between the accelerations caused by the perturbations $\phi_{11}=1$ and $\phi_{21}=1$ depend on the factors $f=(\mathbf{r}\cdot\mathbf{v})^2/(\mathbf{r}\cdot\mathbf{r}\ \mathbf{v}\cdot\mathbf{v})$ and $f^2$. But along the satellite orbits defined in Table~\ref{table:galileo_model_parameters},
the angle between $\mathbf{r}$ and $\mathbf{v}$ is always between $80.8^\circ$ and $99.2^\circ$. Therefore,   
the peak values of $f$ and $f^2$ are only 2,6\% and 0,066\%, respectively. Only for orbits of higher eccentricity these factors become significant.

As our model is based on a point-like Earth, the signals reach the ground station both when the satellite is above the horizon and when it is below the horizon. The latter is not possible in reality. In~the figures below, the regions in which the satellite is under the horizon are therefore shaded, since~comparison with observational data is not possible there.

\subsection{Receive Times}\label{subsec:receive times}
We assume that the satellite transmits signals at a constant rate with respect to its proper time $\tau_t$. The zero points of the time axes are fixed by the assumption that one signal is transmitted at coordinate time $t_0$, corresponding to $\tau_t=0$, and received at the ground station at its proper time $\tau_r=0$. This~applies to both the unperturbed and the perturbed scenarios. With the time axes fixed that way, the~shift of the receive times $\tau_r$ caused by the perturbations is calculated according to
\begin{equation}
\Delta\tau_r:=\tau_r(\phi_n\neq 0)-\tau_r(\phi_n=0).
\end{equation}

\begin{figure}[H]
\centering
\includegraphics[height=\figureHeight]{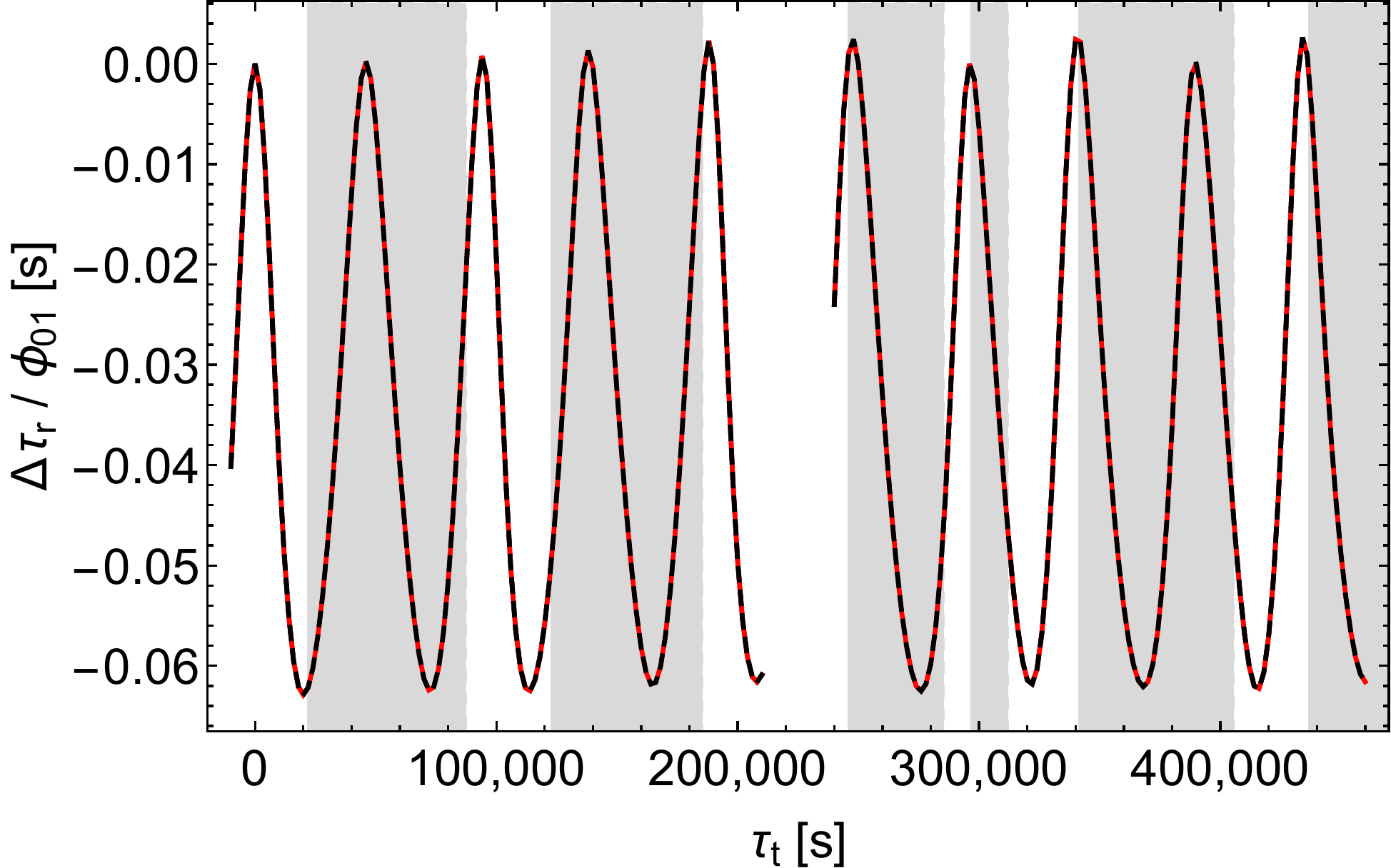}
\caption{Receive time shift $\Delta\tau_r$ divided by $\phi_{01}$. Black dashes $\phi_{01}=10^{-9}$, red dashes $\phi_{01}=10^{-8}$, left~(right) curves $\omega$ = $90^\circ$ ($180^\circ$).}
\label{figure:shift_of_receive_times_phi0_21}
\end{figure}

\begin{figure}[H]
\centering
\includegraphics[height=\figureHeight]{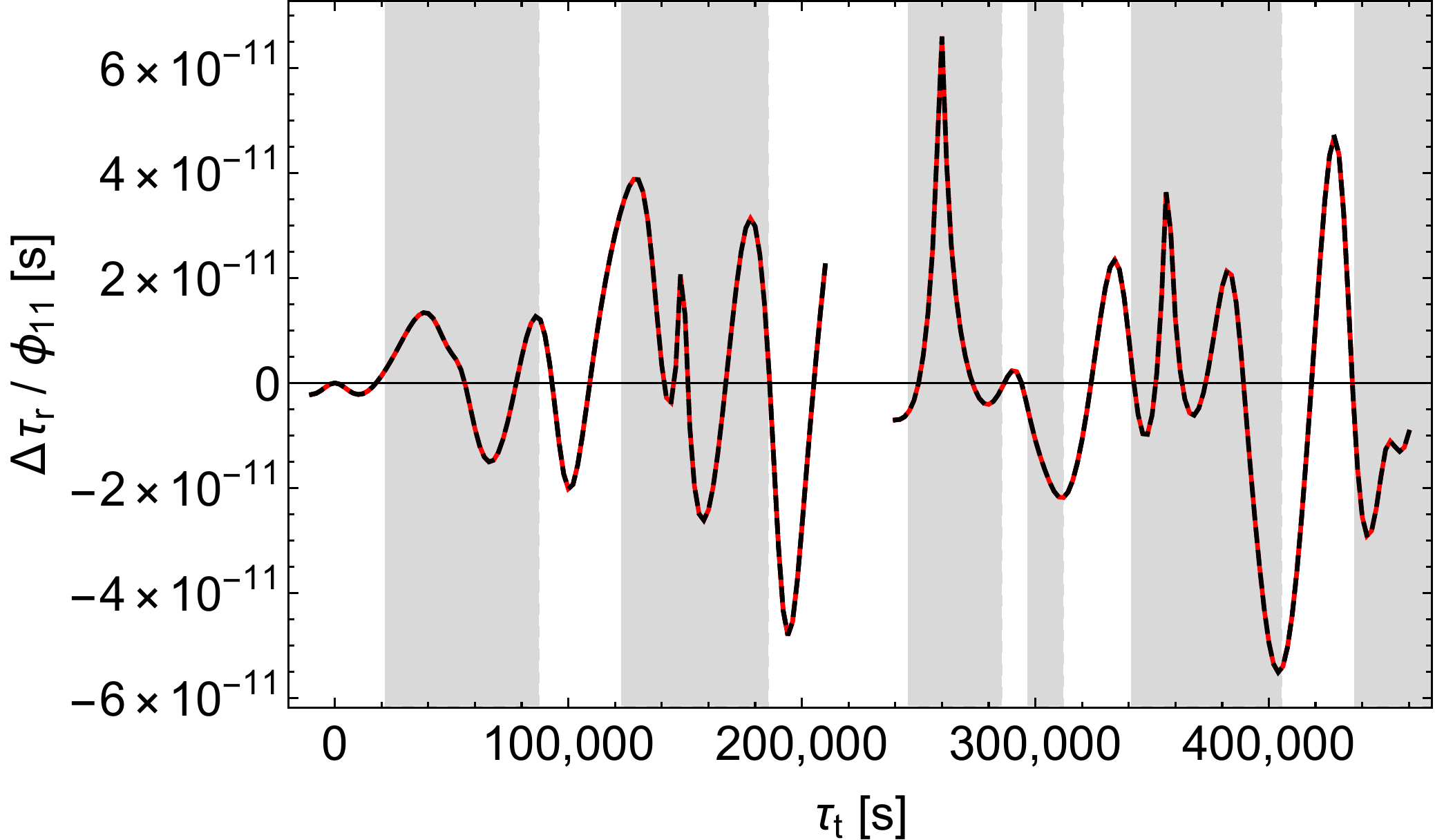}
\caption{{Receive} time shift $\Delta\tau_r$ divided by $\phi_{11}$. Black dashes $\phi_{11}=0.1$, red dashes $\phi_{11}=1$, left~(right) curves $\omega$ = $90^\circ$ ($180^\circ$).}
\label{figure:shift_of_receive_times_phi1_21}
\end{figure}

\begin{figure}[H]
\centering
\includegraphics[height=\figureHeight]{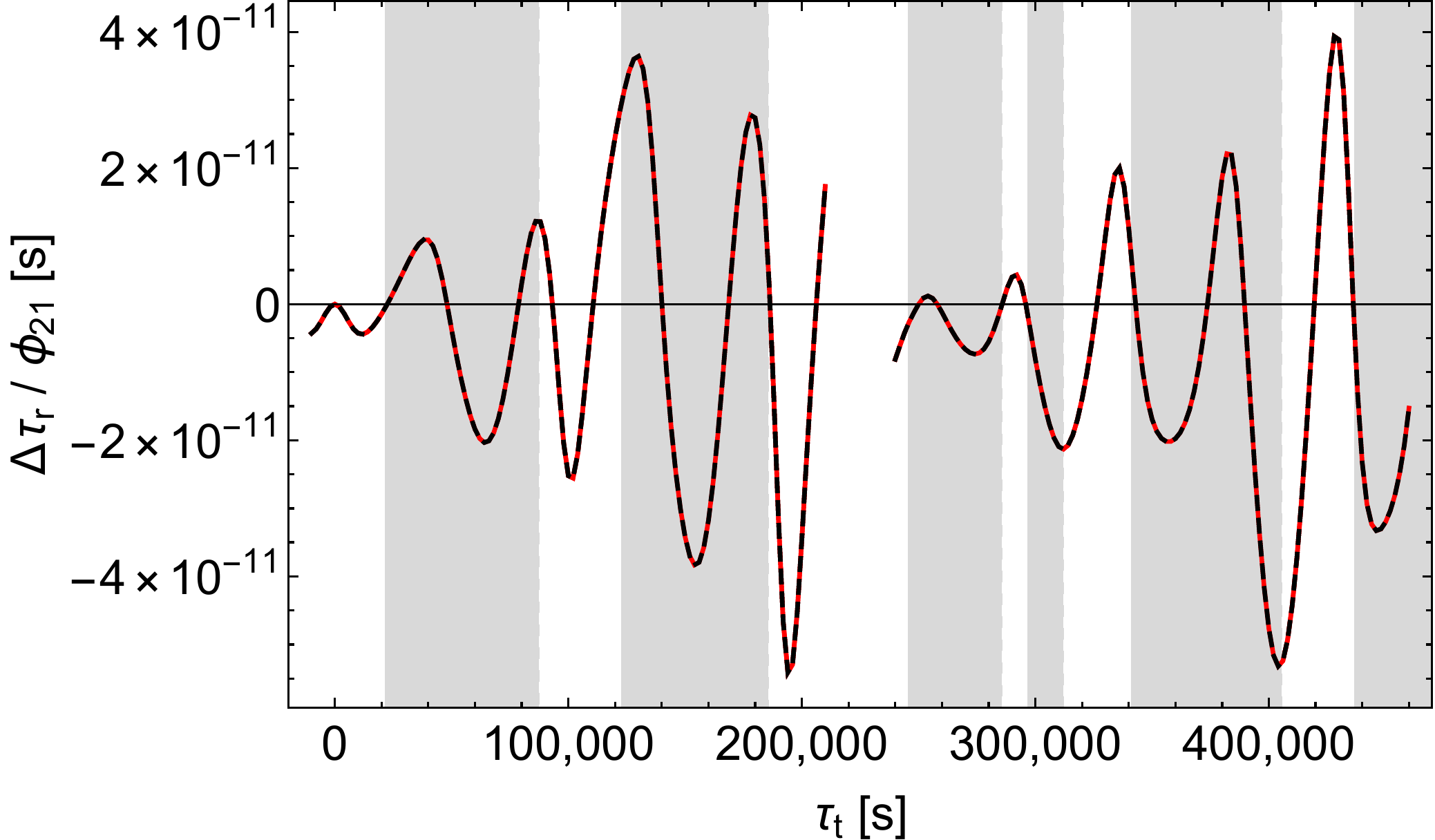}
\caption{Receive time shift $\Delta\tau_r$ divided by $\phi_{21}$. Black dashes $\phi_{21}=0.1$, red dashes $\phi_{21}=1$, left (right) curves $\omega$ = $90^\circ$ ($180^\circ$).}
\label{figure:shift_of_receive_times_phi2_21}
\end{figure}

The results are shown in Figures~\ref{figure:shift_of_receive_times_phi0_21}--\ref{figure:shift_of_receive_times_phi2_21}. We see that $\phi_0$, $\phi_1$, and $\phi_2$ affect the receive times qualitatively different. $\phi_0$ causes a simple and almost periodic pattern with $\Delta\tau_r\ <\ 0$ for the most part, whereas $\phi_1$ and $\phi_2$ cause rich and irregular patterns with varying amplitudes.

Accordingly, $\phi_0$-perturbations can easily be distinguished from $\phi_1$- and $\phi_2$-perturbations by analysing the shifts of the receive times.
But it seems that $\phi_1$- and $\phi_2$-perturbations cannot be easily separated this way, at least not in the case of the satellites Galileo 5 and 6.

\unskip
\unskip

\begin{figure}[H]
\centering
\includegraphics[height=\figureHeight]{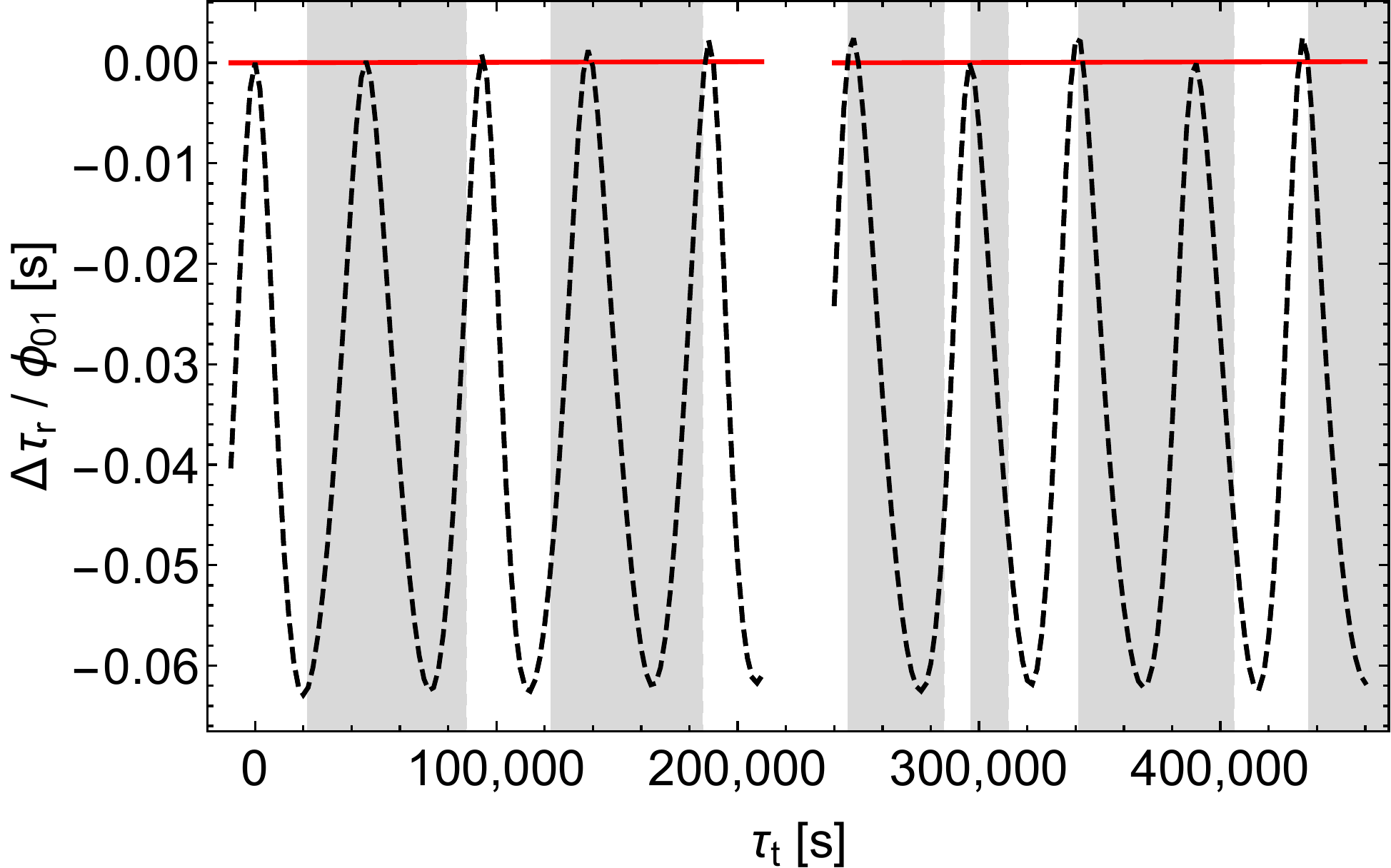}
\caption{Receive time shift $\Delta\tau_r$ divided by $\phi_{01}$. Black dashes $\phi_{01}=10^{-8}$ only for the satellite orbit calculation, red lines $\phi_{01}=10^{-8}$ except for the satellite orbit calculation, left (right) curves \mbox{$\omega$ = $90^\circ$ ($180^\circ$)}.}
\label{figure:shift_of_receive_times_phi0_43}
\end{figure}

\begin{figure}[H]
\centering
\includegraphics[height=\figureHeight]{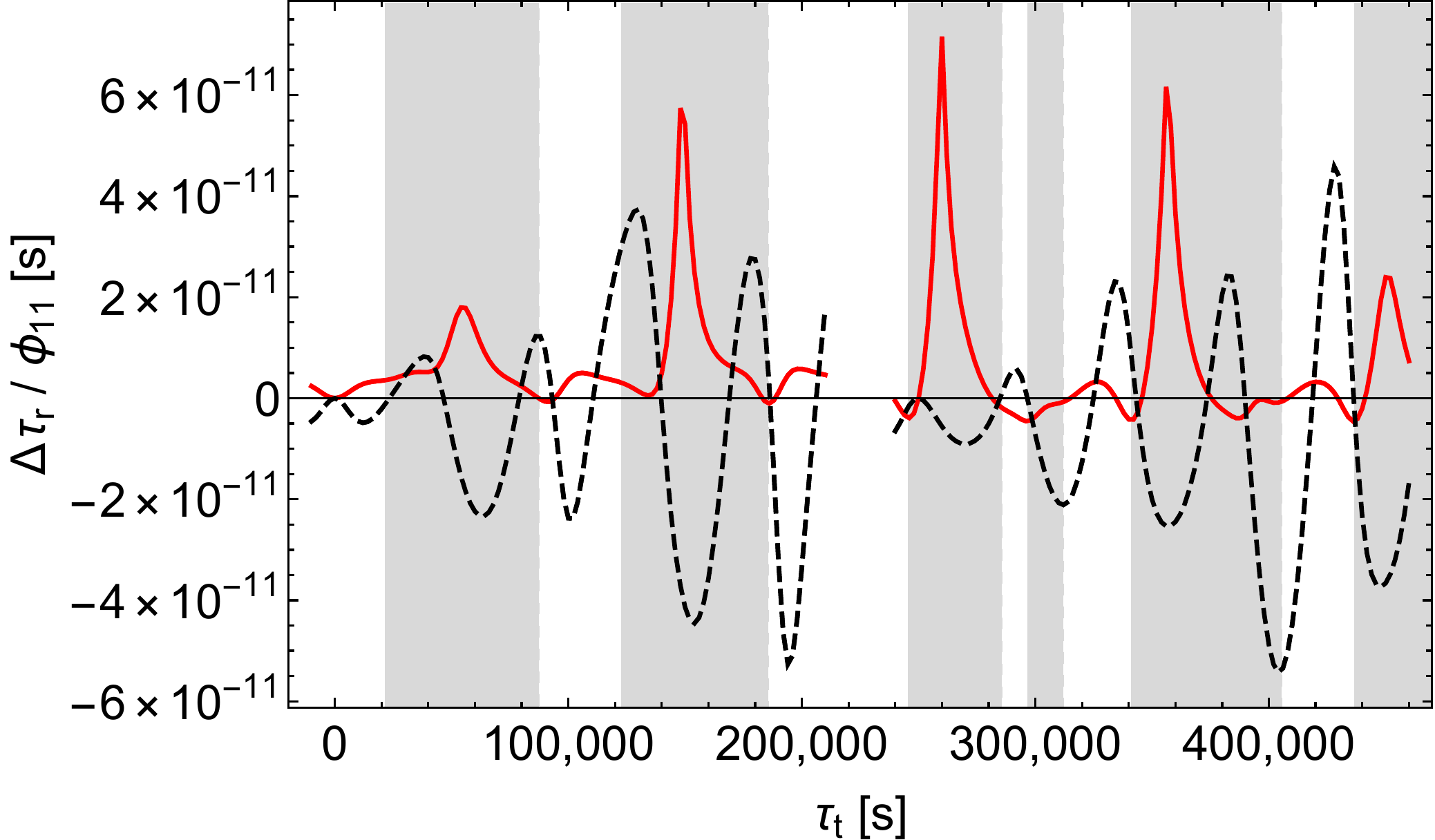}
\caption{Receive time shift $\Delta\tau_r$ divided by $\phi_{11}$. Black dashes $\phi_{11}=1$ only for the satellite orbit calculation, red lines $\phi_{11}=1$ except for the satellite orbit calculation, left (right) curves $\omega$~=~$90^\circ$~($180^\circ$).}
\label{figure:shift_of_receive_times_phi1_43}
\end{figure}

\begin{figure}[H]
\centering
\includegraphics[height=\figureHeight]{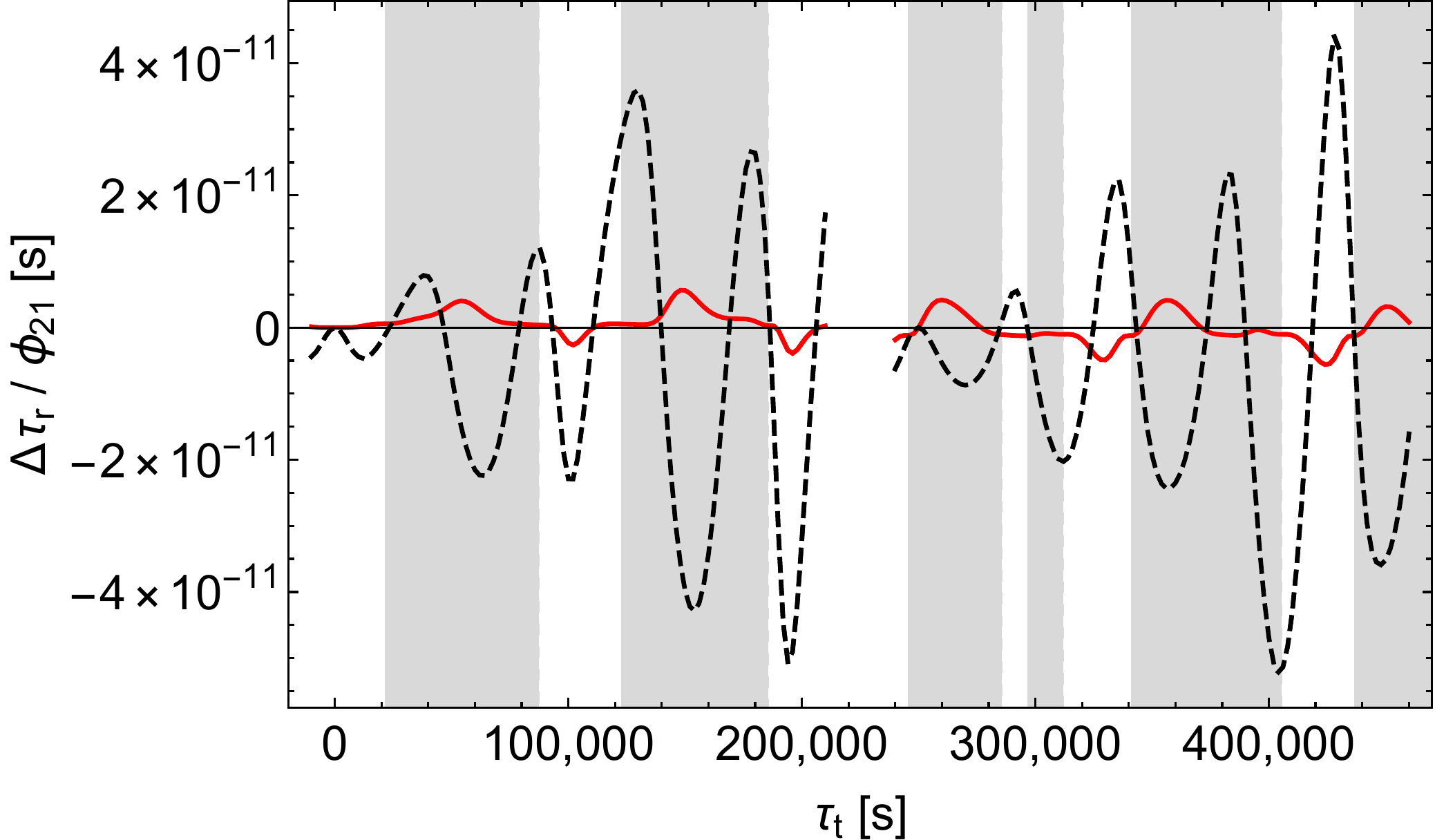}
\caption{Receive time shift $\Delta\tau_r$ divided by $\phi_{21}$. Black dashes $\phi_{21}=1$ only for the satellite orbit calculation, red lines $\phi_{21}=1$ except for the satellite orbit calculation, left (right) curves $\omega$~= $90^\circ$ ($180^\circ$).}
\label{figure:shift_of_receive_times_phi2_43}
\end{figure}

To get more information and to find out how to possibly achieve complete separation, we~calculated the receive time shifts in two hybrid models with the same orbit parameters as above, as~introduced in Section~\ref{sec:hybrid_models}.
In the first model the perturbations affect only the satellite orbit.
In the second model the perturbations affect everything but the satellite orbit, i.e., the signal propagation and the proper times of the satellite and the ground station.
Figures~\ref{figure:shift_of_receive_times_phi0_43}--\ref{figure:shift_of_receive_times_phi2_43} show the results.

\unskip
\unskip

In the case of a pure $\phi_0$-perturbation the receive time shift caused by the perturbation of the orbit (``orbit contribution'') completely dominates the receive time shift caused by the perturbation of the signals and the proper times (``signal contribution''), as can be seen from Figure~\ref{figure:shift_of_receive_times_phi0_43}.
In contrast, in the case of a pure $\phi_1$-perturbation both contributions are of about the same size, see Figure~\ref{figure:shift_of_receive_times_phi1_43}. In addition, there is a characteristic pattern: sharp positive peaks in the signal contribution tend to coincide with negative extrema in the orbit contribution.
Thus, Figure~\ref{figure:shift_of_receive_times_phi1_43} reveals the hidden compensating effects in Figure~\ref{figure:shift_of_receive_times_phi1_21}.
Finally, according to Figure~\ref{figure:shift_of_receive_times_phi2_43}, in the case of a pure $\phi_2$-perturbation the orbit contribution dominates already during the first revolutions of the satellite, although less than with a $\phi_0$-perturbation. However, the receive time shift (red dashes in Figure~\ref{figure:shift_of_receive_times_phi2_21}) is well approximated by the orbit contribution (black dashes in Figure~\ref{figure:shift_of_receive_times_phi2_43}).
It is also notable that the signal contributions of $\phi_1$ and $\phi_2$ tend to have the same sign in the shaded regions and to have opposite signs in the middles of the unshaded regions.

The following considerations make plausible that $\phi_0$, $\phi_1$, and $\phi_2$ give qualitatively different results.

Under the constraint of same sidereal period in the unperturbed and in the perturbed models (cf. Section~\ref{subsec:sidereal period}),
a $\phi_0$-perturbation results in a considerable change of the satellite orbit's semi-major axis (this follows from a simple consideration similar to that in the last paragraph of Section~\ref{sec:LPH12}).
That~change is in turn associated with a considerable shift of the receive times and receive frequencies.
In contrast, $\phi_1$ and $\phi_2$ affect the semi-major axis and hence the orbital motion of the satellite only slightly (cf. Figures~\ref{figure:galileo_shift_satellite_position_0}--\ref{figure:galileo_shift_satellite_position_2} for the effect of $\phi_0$, $\phi_1$, and $\phi_2$ on the satellite orbit).

The relatively sharp and strong peaks in the receive time shift that occur only with $\phi_1$ and only in the shaded regions (one peak every time the satellite is near the nadir) can be understood as the results of fictitious experiments of light deflection at a point mass, and it is already known from Section~\ref{sec:LPH12_light_deflection} that the effect is much stronger with $\phi_1$ than with $\phi_2$.

As mentioned at the end of Section~\ref{sec:Finsler_spacetime}, $\phi_2$ does not affect pure radial velocities.
This makes plausible that the signal contributions of $\phi_1$ and $\phi_2$ (red lines in Figures~\ref{figure:shift_of_receive_times_phi1_43} and~\ref{figure:shift_of_receive_times_phi2_43}) differ significantly also in the middles of the unshaded regions, i.e., where the satellite is near its maximum height above the horizon seen from the ground station and the signal velocity has the largest radial component.

In summary, Figures~\ref{figure:shift_of_receive_times_phi1_43} and~\ref{figure:shift_of_receive_times_phi2_43} demonstrate that in the case of the satellites Galileo 5 and 6 the signal contributions of $\phi_1$ and $\phi_2$ differ more than the orbit contributions. But in the solely accessible unshaded regions the orbit contributions dominate and hamper the reliable separation.
However, in~Section~\ref{sec:high_eccentricity_model} we will demonstrate that for highly eccentric orbits the separation is possible and in this case the different effect of $\phi_1$ and $\phi_2$ on almost radial signals is crucial.
Therefore, one has to conclude that the possibility of the separation of the receive time shifts caused by $\phi_1$ and $\phi_2$ depends on the orbit of the satellite.

\subsection{Receive Frequencies}\label{subsec:receive frequencies}
\begin{figure}[H]
\centering
\includegraphics[height=\figureHeight]{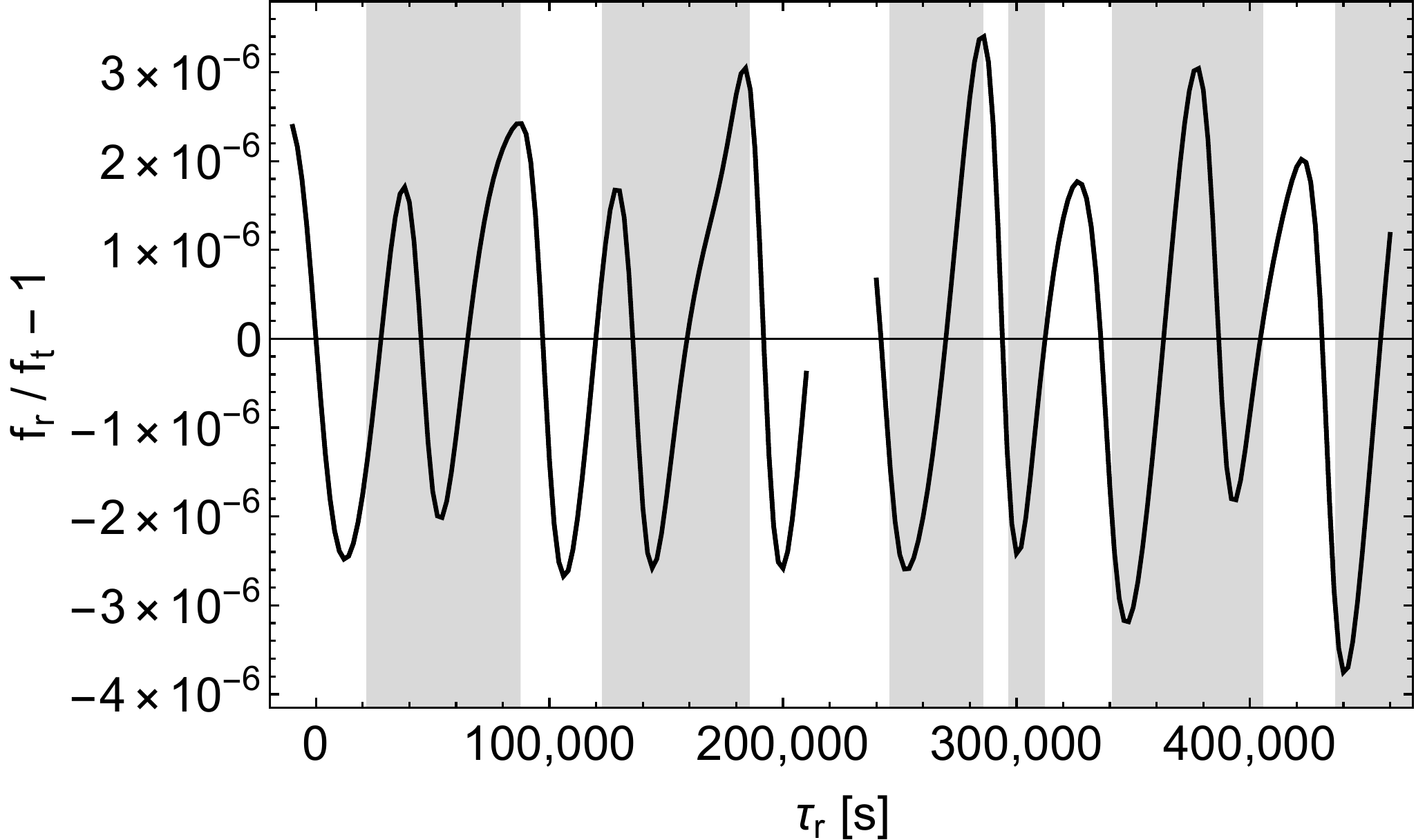}
\caption{Relative frequency shift $f_r$/$f_t-1$ at the ground station in the Schwarzschild spacetime, left~(right) curve $\omega$ = $90^\circ$ ($180^\circ$).}
\label{figure:galileo_relative_frequency_shift_schwarzschild_1}
\end{figure}

Next we consider the receive frequency shift at the ground station caused by $\phi_0$, $\phi_1$, and $\phi_2$. For~comparison, Figure~\ref{figure:galileo_relative_frequency_shift_schwarzschild_1} shows the relative shift between transmit frequency $f_t$ and receive frequency $f_r$ in the Schwarzschild spacetime. In the Schwarzschild spacetime the frequency shift is dominated by the Doppler effect resulting from the relative motion of the satellite and the ground station.

To analyse the effects of the perturbation functions on the frequency shift we use the equation

\begin{equation}\label{equation:relative_receive_frequency_shift}
\frac{\Delta f_r}{f_t}:=\frac{f_r(\phi_n\neq 0)-f_r(\phi_n=0)}{f_t}=\biggl(\frac{f_r(\phi_n\neq 0)}{f_t}-1\biggr)-\biggl(\frac{f_r(\phi_n=0)}{f_t}-1\biggr)=\Delta\biggl(\frac{f_r}{f_t}-1\biggr).
\end{equation}

\begin{figure}[H]
\centering
\includegraphics[height=\figureHeight]{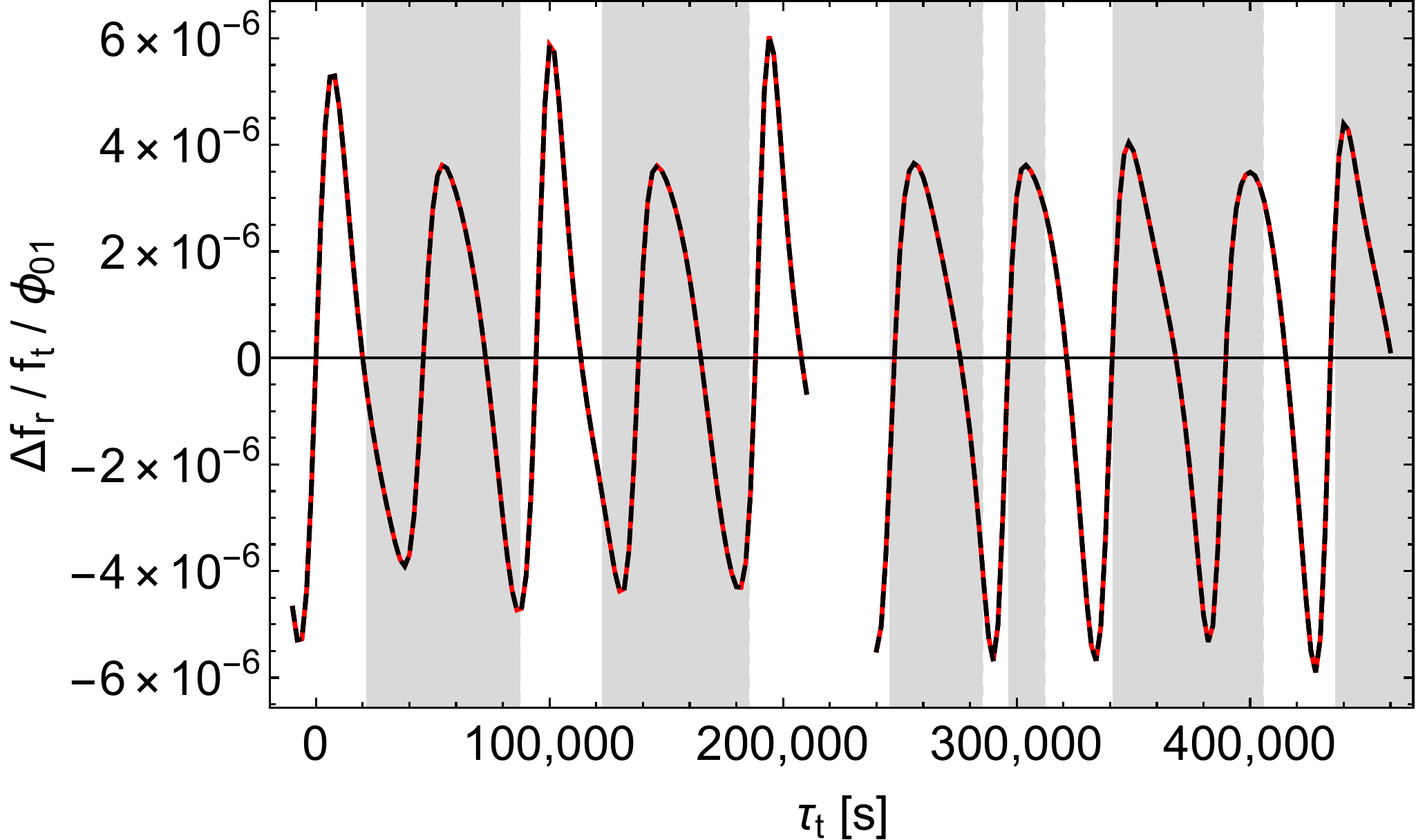}
\caption{Relative receive frequency shift $\Delta f_r/f_t$ divided by $\phi_{01}$. Black dashes $\phi_{01}=10^{-9}$, red dashes $\phi_{01}=10^{-8}$, left (right) curves $\omega$ = $90^\circ$ ($180^\circ$).}
\label{figure:shift_of_receive_frequencies_phi0_21}
\end{figure}

\begin{figure}[H]
\centering
\includegraphics[height=\figureHeight]{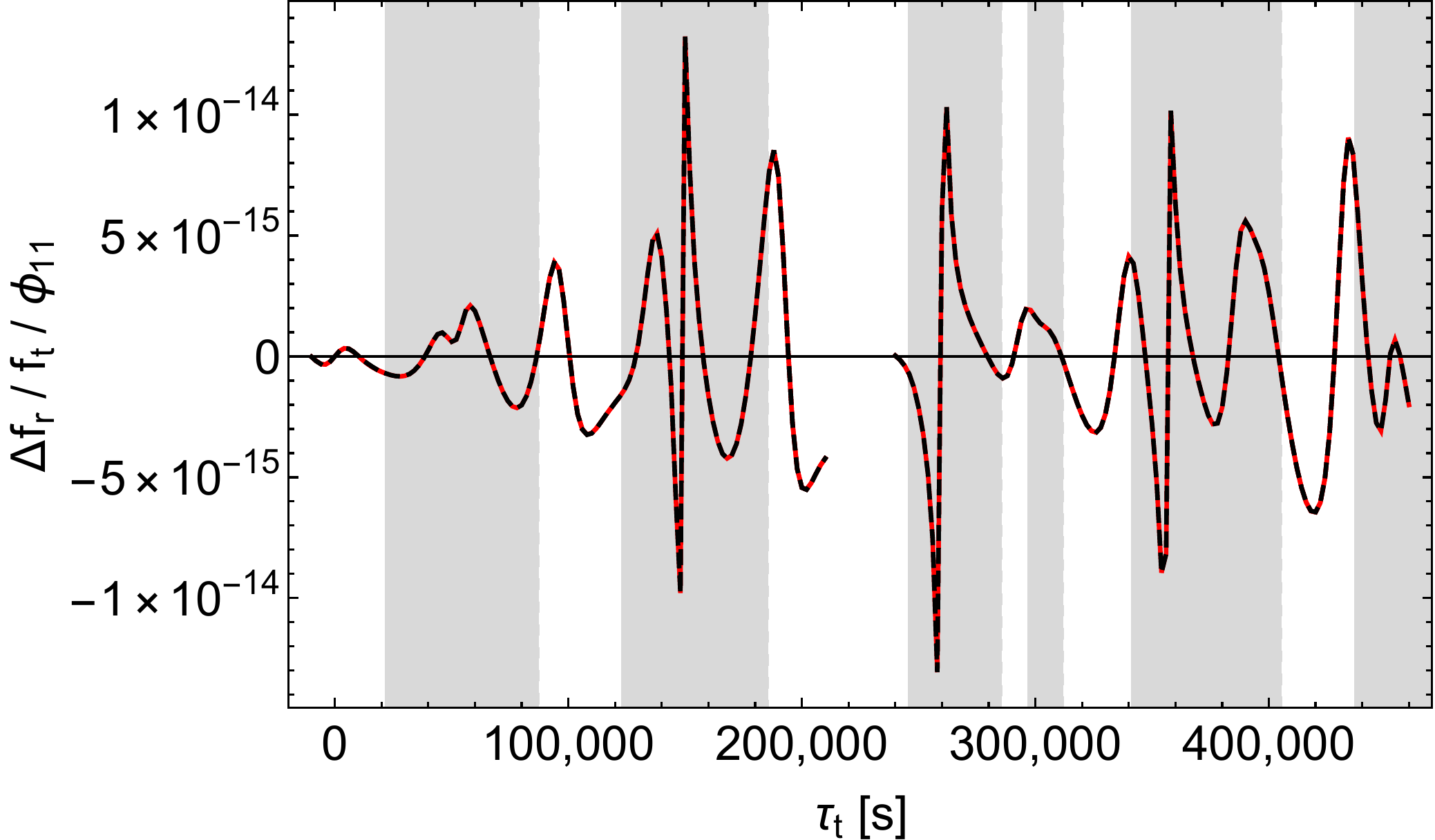}
\caption{Relative receive frequency shift $\Delta f_r/f_t$ divided by $\phi_{11}$. Black dashes $\phi_{11}=0.1$, red dashes $\phi_{11}=1$, left (right) curves $\omega$ = $90^\circ$ ($180^\circ$).}
\label{figure:shift_of_receive_frequencies_phi1_21}
\end{figure}

\begin{figure}[H]
\centering
\includegraphics[height=\figureHeight]{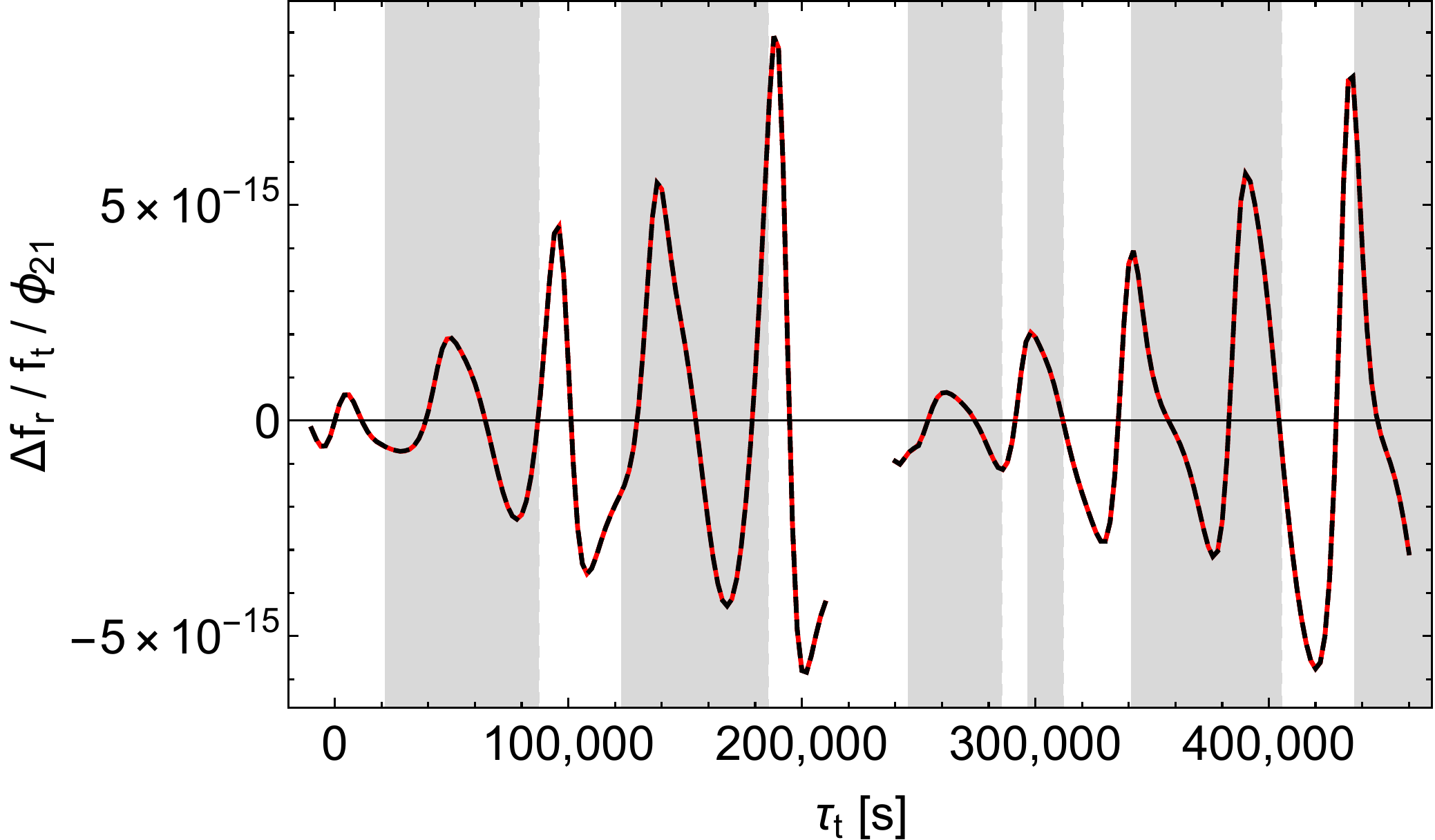}
\caption{Relative receive frequency shift $\Delta f_r/f_t$ divided by $\phi_{21}$. Black dashes $\phi_{21}=0.1$, red dashes $\phi_{21}=1$, left (right) curves $\omega$ = $90^\circ$ ($180^\circ$).}
\label{figure:shift_of_receive_frequencies_phi2_21}
\end{figure}

Since $\Delta f_r/f_t$ as defined by \eqref{equation:relative_receive_frequency_shift} is essentially given by $-d\Delta\tau_r/d\tau_t$, analogous patterns as with $\Delta\tau_r$ are to be expected. Indeed, the graphical representation of our results in Figures~\ref{figure:shift_of_receive_frequencies_phi0_21}--\ref{figure:shift_of_receive_frequencies_phi2_21} confirms this expectation.

\unskip

As with $\Delta\tau_r$, we see that $\phi_0$, $\phi_1$, and $\phi_2$ affect the received signals qualitatively different. Again $\phi_0$ causes a simple, almost periodic pattern with almost constant amplitude, whereas $\phi_1$ and $\phi_2$ cause patterns with varying amplitudes, and again only $\phi_1$ causes sharp peaks in $\Delta f_r/f_t$ which are located in the shaded regions of Figure~\ref{figure:shift_of_receive_frequencies_phi1_21}. However, unlike with $\Delta\tau_r$, these peaks come in pairs oriented downward and upward, as is expected from the explanation following \eqref{equation:relative_receive_frequency_shift}.

\begin{figure}[H]
\centering
\includegraphics[height=\figureHeight]{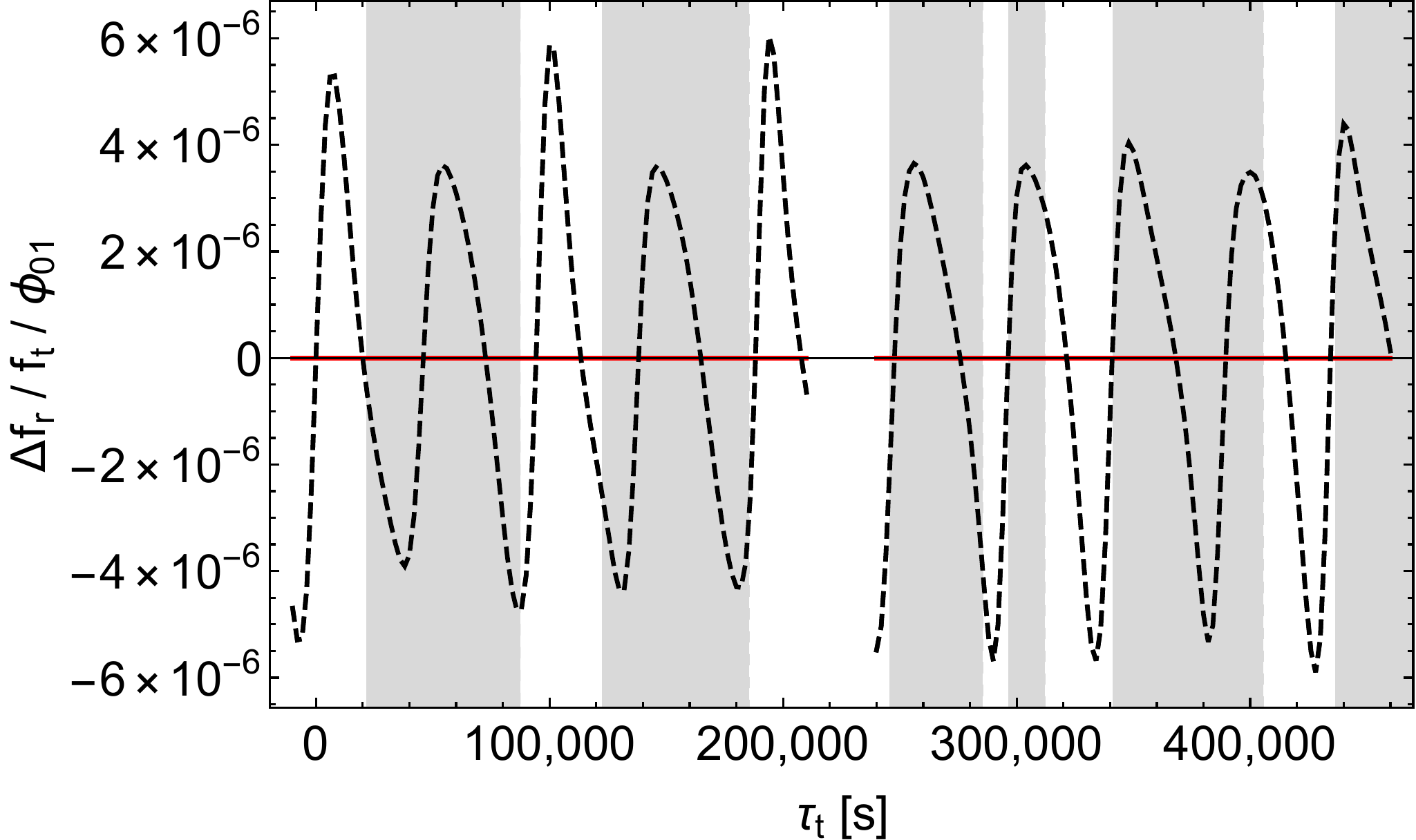}
\caption{Relative receive frequency shift $\Delta f_r/f_t$ divided by $\phi_{01}$. Black dashes $\phi_{01}=10^{-8}$ only for the satellite orbit calculation, red lines $\phi_{01}=10^{-8}$ except for the satellite orbit calculation, left (right) curves $\omega$ = $90^\circ$ ($180^\circ$).}
\label{figure:shift_of_receive_frequencies_phi0_43}
\end{figure}

\begin{figure}[H]
\centering
\includegraphics[height=\figureHeight]{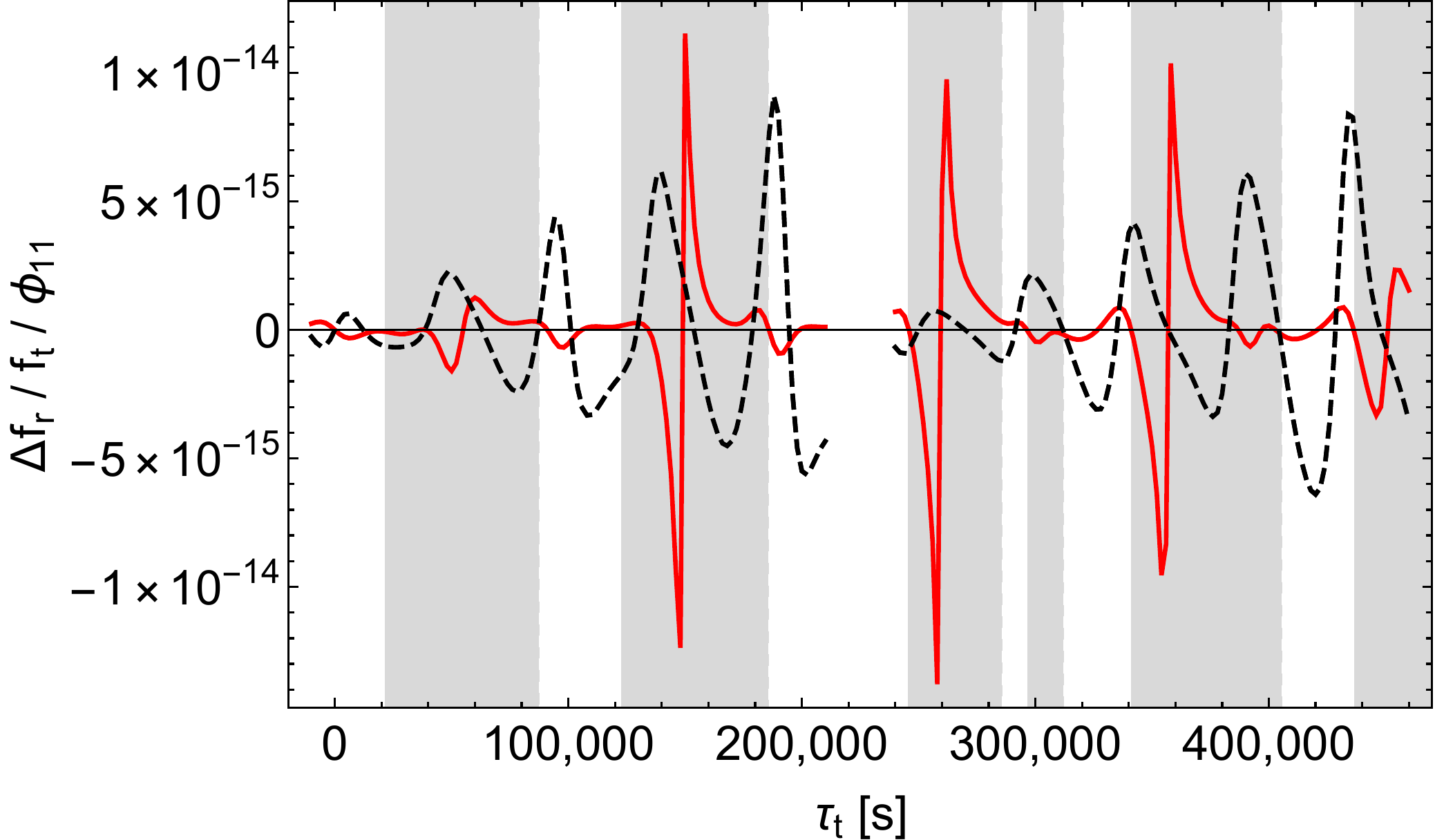}
\caption{Relative receive frequency shift $\Delta f_r/f_t$ divided by $\phi_{11}$. Black dashes $\phi_{11}=1$ only for the satellite orbit calculation, red lines $\phi_{11}=1$ except for the satellite orbit calculation, left (right) curves \mbox{$\omega$ = $90^\circ$ ($180^\circ$).}}
\label{figure:shift_of_receive_frequencies_phi1_43}
\end{figure}

\begin{figure}[H]
\centering
\includegraphics[height=\figureHeight]{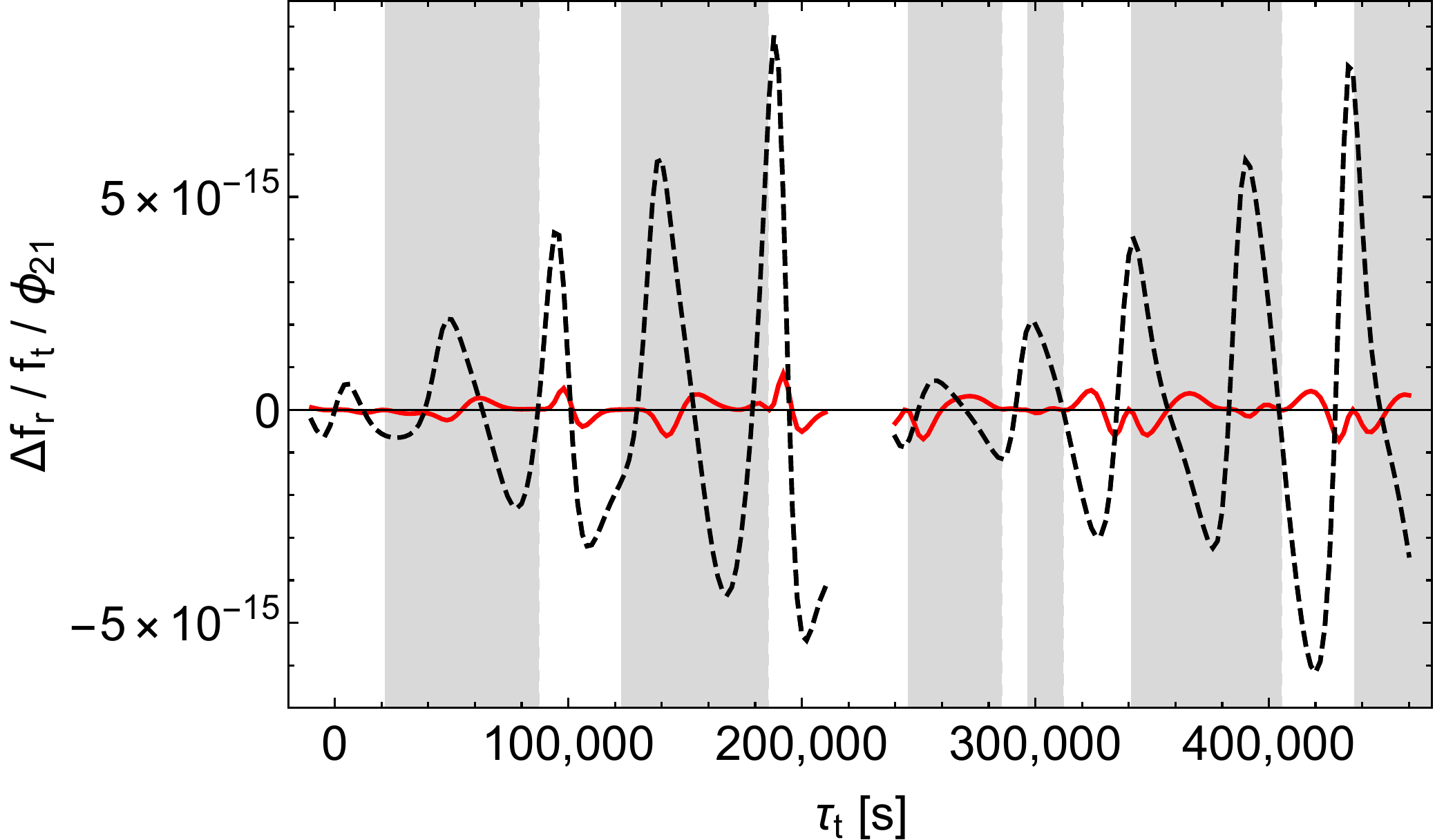}
\caption{Relative receive frequency shift $\Delta f_r/f_t$ divided by $\phi_{21}$. Black dashes $\phi_{21}=1$ only for the satellite orbit calculation, red lines $\phi_{21}=1$ except for the satellite orbit calculation, left (right) curves \mbox{$\omega$ = $90^\circ$ ($180^\circ$).}}
\label{figure:shift_of_receive_frequencies_phi2_43}
\end{figure}

Figures~\ref{figure:shift_of_receive_frequencies_phi0_43}--\ref{figure:shift_of_receive_frequencies_phi2_43} show the corresponding results of our hybrid model calculations.

\unskip
\unskip
\unskip

As with $\Delta\tau_r$, we see that only $\phi_1$ perturbs the signal propagation to an extend that contributes significantly to the whole effect (and again only in the shaded regions).

Finally we observe that $\phi_2$ causes a clearly increasing amplitude of $\Delta f_r/f_t$ (see Figure~\ref{figure:shift_of_receive_frequencies_phi2_21}) and that $\phi_1$ should also cause this effect. Namely, the black dashes in Figure~\ref{figure:shift_of_receive_frequencies_phi1_43} indicate that the ``orbit contribution'' of $\Delta f_r/f_t$ also increases in time and will eventually dominate the non-increasing (or at least much less increasing) ``signal contribution''. Simulations of $\Delta\tau_r$ and $\Delta f_r/f_t$ over time spans of up to $10^7$ seconds confirm this expectation (and show no new features).

As expected, for the frequency shifts the situation regarding the desired separation of the effects of $\phi_1$ and $\phi_2$ is very similar to that for the receive time shifts.

In summary we have to state that at least in the examples considered here, based on the Galileo 5 and 6 satellites, the effects of a genuine Finslerian ($\phi_2$-) perturbation and of a $\phi_1$-perturbation can hardly be separated. This assessment takes into account that noise and additional gravitational or non-gravitational perturbations may cover or distort the specific structures in the signals.


\section{High-Eccentricity Orbit Simulation}\label{sec:high_eccentricity_model}

This section deals with a model satellite in a prograde equatorial Earth orbit with perigee distance $\rho_p$~=~23,500 km and perigee speed $v_p=5.370$ km/s and thus apogee distance $\rho_a$~=~133,227 km, semi-major axis $a$~=~78,364 km, and high eccentricity $e=0.7001$.
The ground station is on the equator and we assume $\varphi_{station} = \varphi_{satellite} + \pi/2$ at $t=0$.

This choice of parameter values is justified as follows. On the one hand, the perigee distance should not be significantly smaller than in the case of the Galileo 5 and 6 satellites since the satellite should not enter the inner Van Allen radiation belt. On the other hand, our analytical calculations (these are too extensive to be presented here) show that at an angle $\alpha$ (describing the direction of the satellite's velocity and defined by Equation \eqref{alpha_definition}) of about 0.86 the absolute value of the $\alpha$-dependent factor of the $\phi_2$-part of $\ddot{\varphi}$ is maximal, see Equation \eqref{ddot_varphi} (whereas $\phi_1$ does not enter the expression for $\ddot{\varphi}$). It turns out that for $\alpha$ to be about 0.86 somewhere on the trajectory, the eccentricity must be at least about 0.7. This explains the relatively great apogee distance. (When comparing the following figures with the respective figures in the previous section please note that now the semi-major axis is larger than the radius of the geostationary orbit. Seen from the ground station the high eccentricity satellite is thus moving from east to west whereas the Galileo 5 and 6 satellites are moving from west to east).

As in Section~\ref{sec:Galileo} the initial values of the perigee speed are so adjusted that the sidereal period is the same with and without perturbation (absolute value of relative shift $<$$10^{-17}$).

Since with regard to $\phi_0$ there are no substantially different patterns than in Section~\ref{sec:Galileo}, we will discuss here only the disentanglement of the effects of $\phi_1$ and $\phi_2$.

\begin{figure}[H]
\centering
\includegraphics[height=\figureHeight]{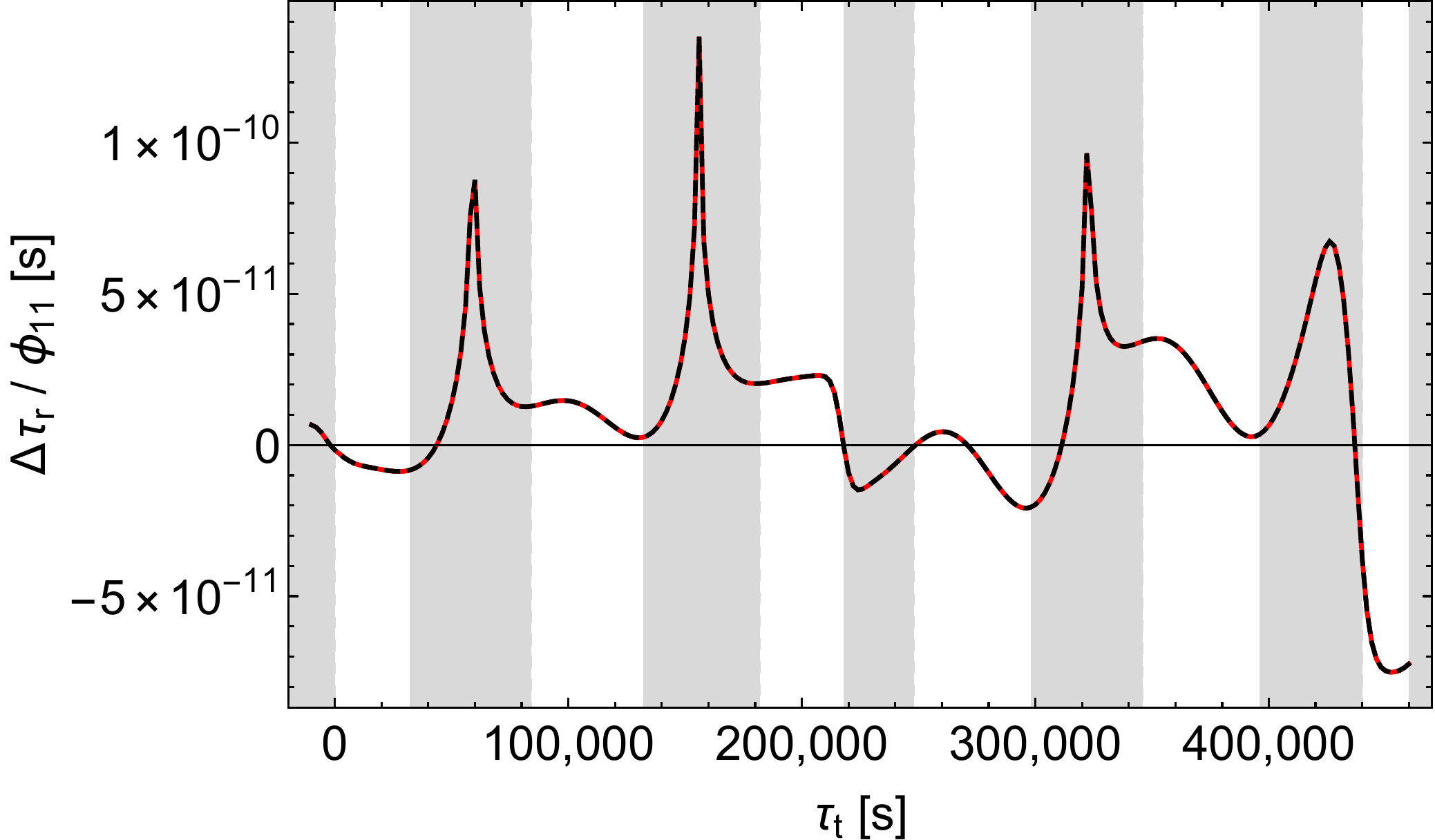}
\caption{Receive time shift $\Delta\tau_r$ divided by $\phi_{11}$. Black dashes $\phi_{11}=0.1$, red dashes $\phi_{11}=1$, $e=0.7001$.}
\label{figure:high_eccentricity_shift_of_receive_times_phi1_21}
\end{figure}

\begin{figure}[H]
\centering
\includegraphics[height=\figureHeight]{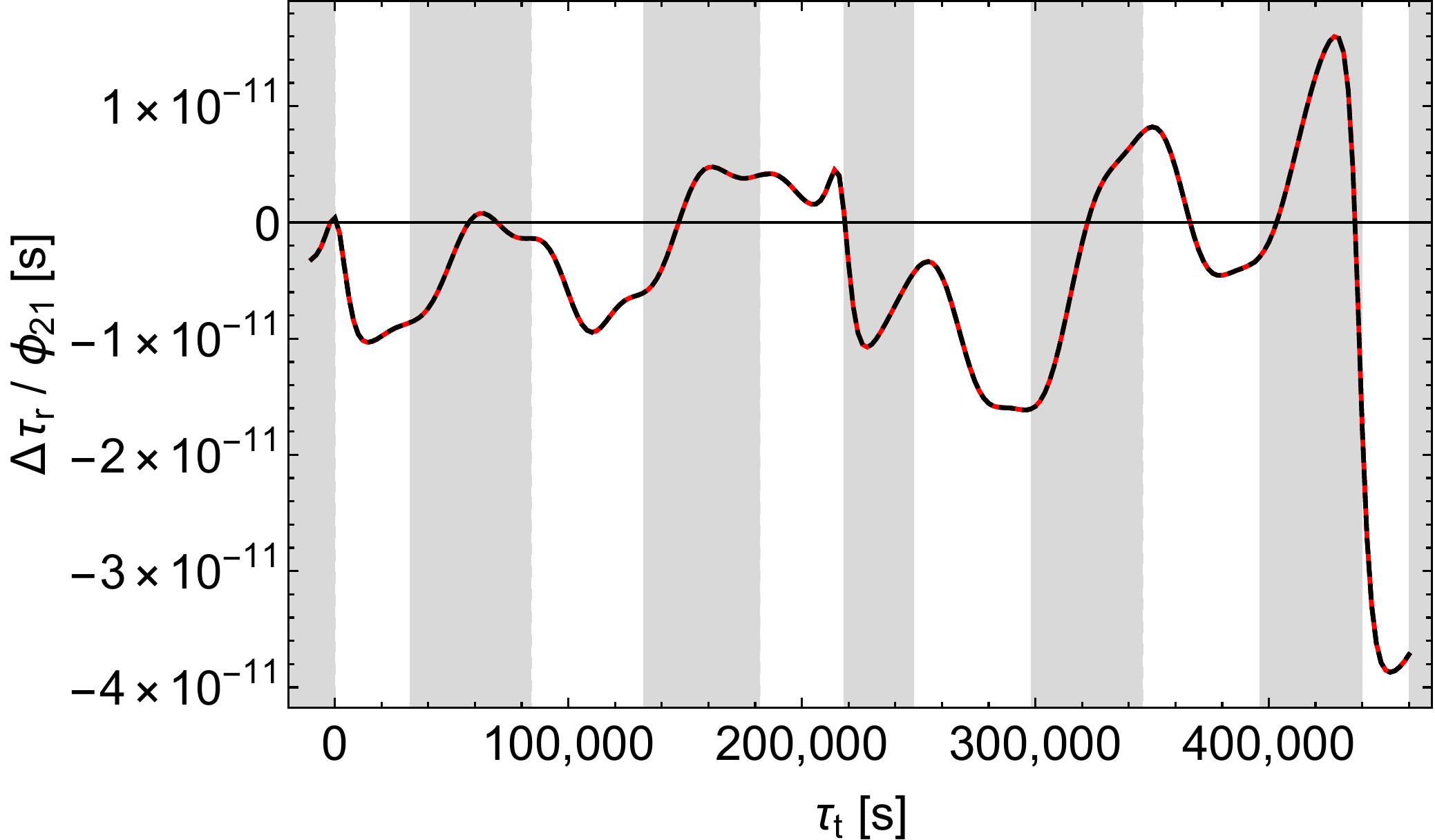}
\caption{Receive time shift $\Delta\tau_r$ divided by $\phi_{21}$. Black dashes $\phi_{21}=0.1$, red dashes $\phi_{21}=1$, $e=0.7001$.}
\label{figure:high_eccentricity_shift_of_receive_times_phi2_21}
\end{figure}

Figures~\ref{figure:high_eccentricity_shift_of_receive_times_phi1_21} and~\ref{figure:high_eccentricity_shift_of_receive_times_phi2_21} show that the effects of $\phi_1$ and $\phi_2$ on the receive times are significantly different, even in the unshaded regions (please note the different scaling of the ordinate axes). Especially, unlike with the Galileo satellites, there are periods of time in which $\phi_1$ and $\phi_2$ shift the receive times in opposite directions.

\unskip

\begin{figure}[H]
\centering
\includegraphics[height=\figureHeight]{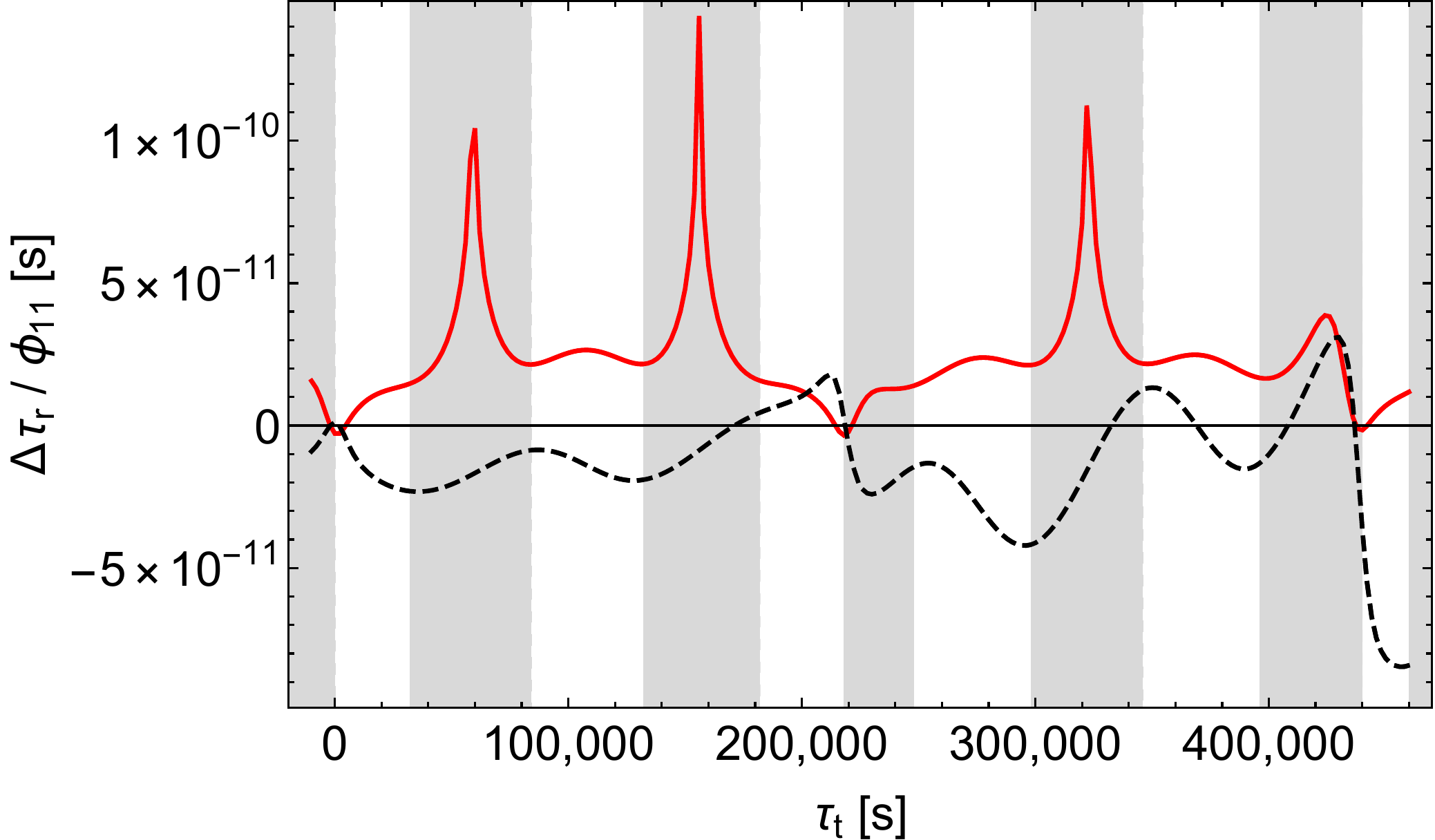}
\caption{Receive time shift $\Delta\tau_r$ divided by $\phi_{11}$. Black dashes $\phi_{11}=1$ only for the satellite orbit calculation, red lines $\phi_{11}=1$ except for the satellite orbit calculation, $e=0.7001$.}
\label{figure:high_eccentricity_shift_of_receive_times_phi1_43}
\end{figure}

\begin{figure}[H]
\centering
\includegraphics[height=\figureHeight]{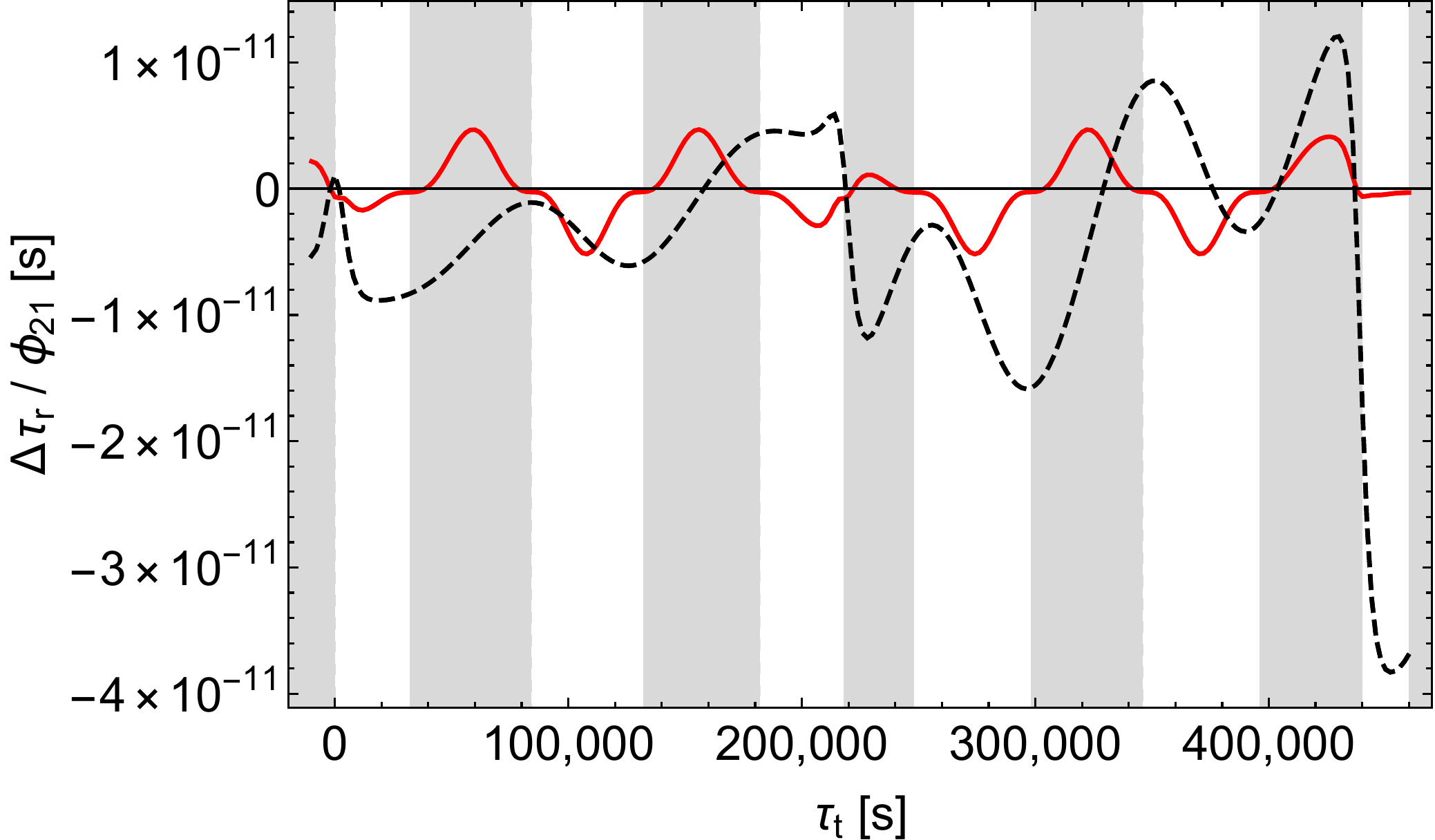}
\caption{Receive time shift $\Delta\tau_r$ divided by $\phi_{21}$. Black dashes $\phi_{21}=1$ only for the satellite orbit calculation, red lines $\phi_{21}=1$ except for the satellite orbit calculation, $e=0.7001$.}
\label{figure:high_eccentricity_shift_of_receive_times_phi2_43}
\end{figure}

Remarkably, in Figures~\ref{figure:high_eccentricity_shift_of_receive_times_phi1_43} and~\ref{figure:high_eccentricity_shift_of_receive_times_phi2_43} the ``orbit'' and the ``signal'' contributions to the shifts calculated with the hybrid models are of the same order of magnitude (whereas with the Galileo satellites the orbit contributions clearly dominate, except in the shaded regions in the case of a $\phi_1$-perturbation).
It is also noteworthy that the mentioned shifts in opposite directions present in some periods of time are solely caused by the signal contributions
because again, as with the Galileo satellites, the orbit contributions of $\phi_1$ and $\phi_2$ to the shifts run in parallel (but unlike with these satellites the orbit contribution of $\phi_2$ amounts to only about half the orbit contribution of $\phi_1$).

\begin{figure}[H]
\centering
\includegraphics[height=\figureHeight]{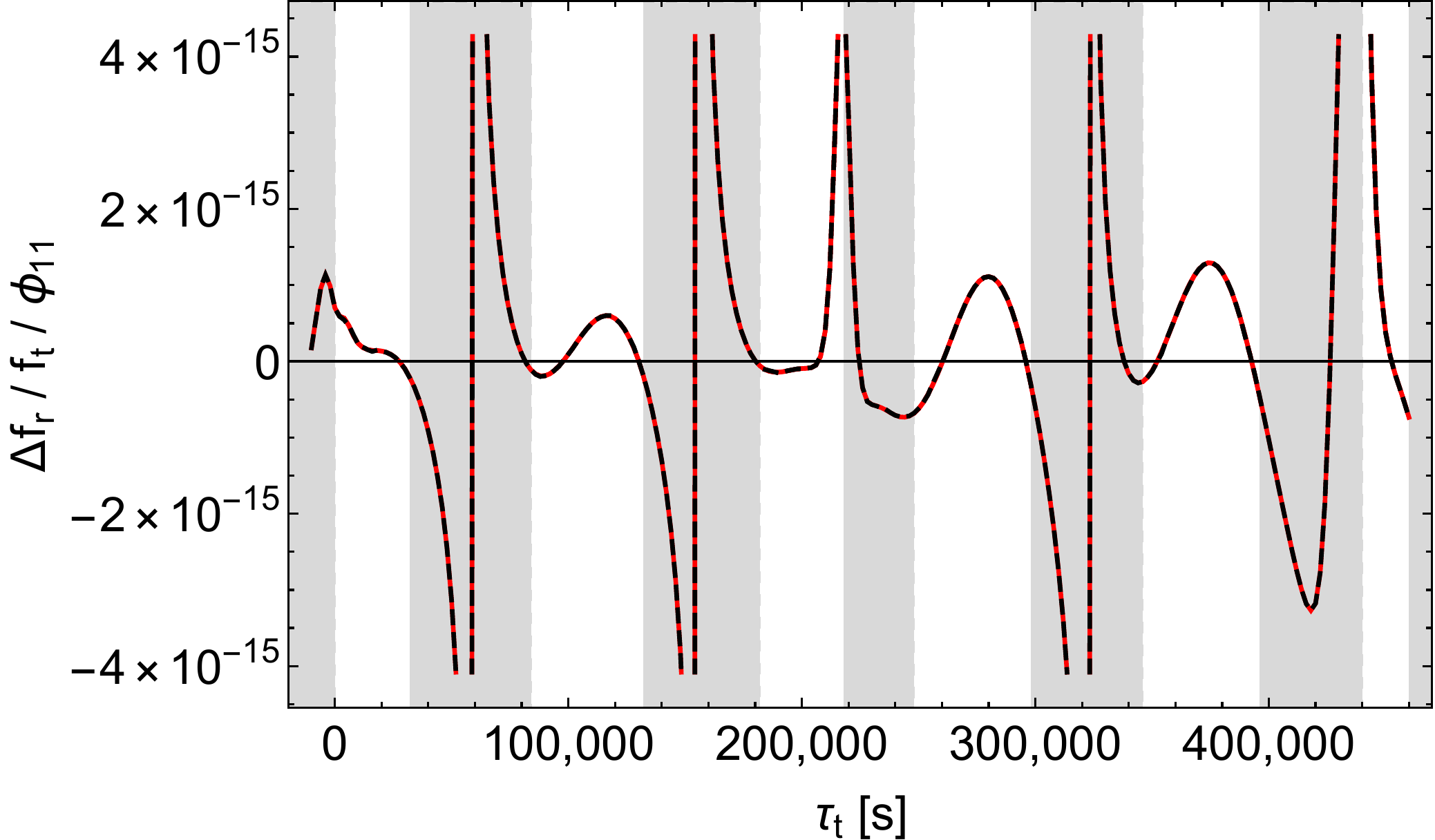}
\caption{Relative receive frequency shift $\Delta f_r/f_t$ divided by $\phi_{11}$. Black dashes $\phi_{11}=0.1$, red dashes $\phi_{11}=1$, $e=0.7001$.}
\label{figure:high_eccentricity_shift_of_receive_frequencies_phi1_21_automatic}
\end{figure}

\begin{figure}[H]
\centering
\includegraphics[height=\figureHeight]{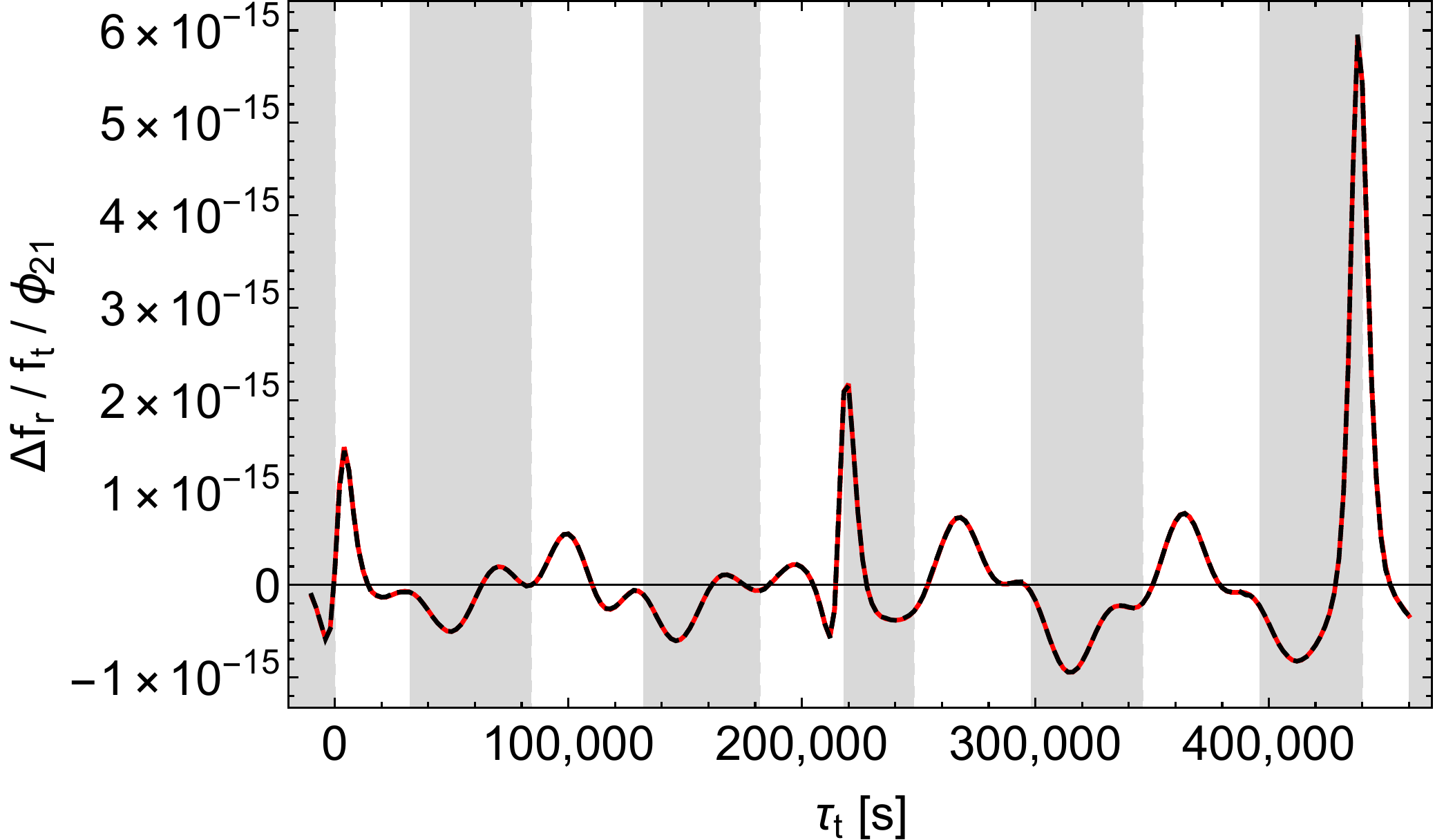}
\caption{Relative receive frequency shift $\Delta f_r/f_t$ divided by $\phi_{21}$. Black dashes $\phi_{21}=0.1$, red dashes $\phi_{21}=1$, $e=0.7001$.}
\label{figure:high_eccentricity_shift_of_receive_frequencies_phi2_21}
\end{figure}

As mentioned in Section~\ref{subsec:receive frequencies}, $\Delta f_r/f_t$ as defined by \eqref{equation:relative_receive_frequency_shift} is essentially given by $-d\Delta\tau_r/d\tau_t$. Accordingly, we expect $\phi_1$ and $\phi_2$ to cause distinguishable structures also in $\Delta f_r/f_t$. Figures~\ref{figure:high_eccentricity_shift_of_receive_frequencies_phi1_21_automatic} and~\ref{figure:high_eccentricity_shift_of_receive_frequencies_phi2_21} confirm our expectation, although the differences are slightly less obvious as with $\Delta\tau_r$.

\unskip
\unskip
\unskip

\section{Conclusions}\label{sec:conclusions}

While purely analytical methods quickly reach their limits in the treatment of the problems considered in this paper,
as can be seen from Section IV of the recent article of Hasse and Perlick~\cite{reference:Hasse_2019},
it is demonstrated here that valuable information about Finslerian effects can be obtained by means of numerical methods, even with today's standard computing systems.
Assuming first order perturbations $\phi_n(\rho)=\phi_{n1}\ r_S/\rho\ (n = 0,\ 1,\ 2)$ in the considered class of spherical symmetric Finsler spacetimes, our numerical simulations in combination with our hybrid models allow us to draw conclusions about the optimal types of orbits for testing models of Finsler gravity.

As can be seen in Section~\ref{sec:Galileo}, the most difficult task is to separate the effects of $\phi_1$ and $\phi_2$.
In~the case of the Galileo 5 and 6 satellites the ``orbit contributions'' to the receive time and frequency shifts caused by $\phi_1$ and $\phi_2$ are practically indistinguishable and dominate the ``signal contributions'' when the satellites are above the horizon, even within the first revolution of the satellite. The ``signal contribution'' is significant only when the satellites are below the horizon {(i.e., in situations of fictitious experiments of light deflection at a point mass)} and only with $\phi_1$.

However, for highly eccentric orbits genuine Finslerian effects can be separated from effects of perturbations of the Schwarzschild spacetime within the Lorentzian geometry.
This is one of the main results and important for the planning of possible future dedicated missions.
Although in the illustrating example of Section~\ref{sec:high_eccentricity_model} the ``orbit contributions'' to the receive time and frequency shifts caused by $\phi_1$ and $\phi_2$ run in parallel too (but with different heights, unlike with the Galileo satellites), the~superimposed ``signal contributions'' differ significantly even when the satellite is above the horizon and make both effects separable.

Unfortunately, the orbit eccentricity of the Galileo 5 and 6 satellites is too small to allow this separation (see our conclusions at the end of Sections~\ref{subsec:receive times} and~\ref{subsec:receive frequencies}).
Nevertheless, these two already available and appropriately equipped satellites in principle allow the determination of the sum of $\phi_{11}$ and $\phi_{21}$ as a valuable first step.
Although, as stated above, along highly eccentric orbits the contributions of $\phi_{1}$ and $\phi_{2}$ to the receive time and frequency shifts differ to an extent that $\phi_{11}$ and $\phi_{21}$ can be separately determined, it helps with the analysis of such an experiment if the sum of $\phi_{11}$ and $\phi_{21}$ is  additionally known from another experiment (the Galileo satellites).

We like to conclude with a few remarks that go beyond the work presented here.

In principle, non-gravitational perturbations (see our remarks in Section~\ref{sec:intro}) may be present that are unknown or insufficiently understood and therefore cannot be modelled.
Nevertheless, one can assume that such perturbations cause no ``Finsler-like'' effects in the receive times and frequencies of the signals.
The reason is that the various (non-Finslerian) perturbations typically show either a nearly diurnal or semidiurnal variation (if they are not of terrestrial origin, e.g., if they are caused by the sun or the moon) or a variation with a leading period that corresponds to the satellite's orbital period (if they are of terrestrial origin).
In the latter case, it is not expected that minima of the non-Finslerian perturbations occur just in the apsides, especially in the perigee, unlike with Finslerian ($\phi_2$-) perturbations, regarded as additional ``forces'', that are very weak in the apsides and strongest somewhere between the apsides.

In view of tests of Finsler gravity, an obvious idea is to extend our approach, in particular the method for determining the signals between satellite and ground station and the concept of hybrid models, to the analysis of other situations of communication.
So one could first think of a two-way time transfer procedure, i.e., signals that travel both ways between two clocks that are being compared.

Finally, the methods presented here may also be adapted to similar tests of other alternatives to general relativity.
In principle, this should be possible at least for theories with well-defined second-order equations of motion for spinless test bodies and light signals in which one can consider a parametrized class of models that includes the Schwarzschild spacetime or other suitable reference solutions of Einstein's field equations.
Examples are the Brans–Dicke theory, Einstein-Cartan models of gravity, and Weyl manifolds as spacetime models.

\vspace{6pt} 

\noindent\textbf{Author Contributions: }
Conceptualization, all authors;
software, I.A. (Sections~\ref{sec:LPH12},~\ref{sec:Galileo} and~\ref{sec:high_eccentricity_model}) and M.P. (Sections~\ref{sec:LPH12_light_deflection} and~\ref{sec:LPH12_perihelion_precession});
formal analysis, I.A. and W.H.;
writing--original draft preparation, I.A. and W.H.;
visualization, I.A.

\appendix

\section{Equations of Motion in Schwarzschild Coordinates in the Plane $ \vartheta= \pi/2$}\label{app:2d_equations}

In this section we briefly describe the derivation of the second order equations of motion in Schwarzschild coordinates in the plane $\vartheta=\pi/2$, with the coordinate time as the independent variable. {The first steps are the elimination of the affine parameter by means of the constants of motion $\mathcal{L}$, $E=-\partial\mathcal{L}/\partial\dot{t}$, and $L=\partial\mathcal{L}/\partial\dot{\varphi}$ (cf. Equations (16)\textendash(19) in~\cite{reference:Laemmerzahl_2012}) and the linearisation of the resulting equations with respect to the perturbation functions. As we are only interested in causal geodesics, we~will only consider geodesics with $\dot{t}\neq0$.}

{The elimination of the affine parameter is based on the constant of motion}
\begin{equation}\label{energy_1_equatorial_plane}
{E:=-\frac{\partial\mathcal{L}}{\partial\dot{t}}=-(1+\phi_0)h_{tt}\dot{t}}
\end{equation}
{and the resulting identity}
\begin{equation}\label{energy_identity}
{-\dfrac{E}{h_{tt}}\dfrac{1}{1+\phi_0}\frac{1}{\dot{t}}=1.}
\end{equation}

{Restricted to the plane $\vartheta=\pi/2$ the Lagrangian \eqref{Lagrangian_A} reads}
\begin{equation}\label{Lagrangian_A_equatorial_plane}
{
2\mathcal{L}=
(1+\phi_0)h_{tt}\dot{t}^2
+(1+\phi_1)h_{rr}\dot{r}^2
+r^2\dot{\varphi}^2
+\phi_2\frac{h_{rr}\dot{r}^2r^2\dot{\varphi}^2}
{h_{rr}\dot{r}^2+r^2\dot{\varphi}^2},
}
\end{equation}
{and the affine parameter can be eliminated by multiplying the right side of \eqref{Lagrangian_A_equatorial_plane} with the square of the left side of \eqref{energy_identity}. This leads to}\begingroup\makeatletter\def\f@size{9.5}\check@mathfonts
\def\maketag@@@#1{\hbox{\m@th\normalsize\normalfont#1}}%
\begin{equation}\label{Lagrangian_A_equatorial_plane_without_affine_parameter}
{
2\mathcal{L}=\dfrac{E^2}{h_{tt}^2}\dfrac{1}{(1+\phi_0)^2}\biggl\{(1+\phi_0)h_{tt}
+(1+\phi_1)h_{rr}\biggl(\dfrac{dr}{dt}\biggr)^2
+r^2\biggl(\dfrac{d\varphi}{dt}\biggr)^2
+\phi_2\dfrac{h_{rr}\biggl(\dfrac{dr}{dt}\biggr)^2r^2\biggl(\dfrac{d\varphi}{dt}\biggr)^2}
{h_{rr}\biggl(\dfrac{dr}{dt}\biggr)^2+r^2\biggl(\dfrac{d\varphi}{dt}\biggr)^2}\biggr\}.
}
\end{equation}\endgroup

{Linearising \eqref{Lagrangian_A_equatorial_plane_without_affine_parameter} with respect to the perturbation functions (especially, replacing $(1+\phi_0)^{-2}$ with its linear approximation $1 - 2\phi_0$) then results in}
\begin{equation}\label{conserved_quantity_1}
\begin{split}
2\mathcal{L} = &\dfrac{E^2}{h_{tt}^2} \biggl\{h_{tt} + h_{rr}\biggl(\dfrac{dr}{dt}\biggr)^2 + r^2\biggl(\dfrac{d\varphi}{dt}\biggr)^2
-\phi_0\biggl(h_{tt} + 2h_{rr}\biggl(\dfrac{dr}{dt}\biggr)^2 + 2r^2\biggl(\dfrac{d\varphi}{dt}\biggr)^2\biggr)\\
&+\phi_1h_{rr}\biggl(\dfrac{dr}{dt}\biggr)^2 + \phi_2\dfrac{h_{rr} \biggl(\dfrac{dr}{dt}\biggr)^2 {r^2} \biggl(\dfrac{d\varphi}{dt}\biggr)^2}{h_{rr}\biggl(\dfrac{dr}{dt}\biggr)^2 + r^2\biggl(\dfrac{d\varphi}{dt}\biggr)^2}
\biggr\}{.}
\end{split}
\end{equation}

{Restricted to the plane $\vartheta=\pi/2$ the constant of motion $L=\partial\mathcal{L}/\partial\dot{\varphi}$ reads}
\begin{equation}\label{angular_momentum}
{
L:=\dfrac{\partial\mathcal{L}}{\partial\dot{\varphi}}
=r^2\dot{\varphi} \biggl(1+\phi_2\dfrac{h_{rr}^2\dot{r}^4}{(h_{rr}\dot{r}^2+r^2\dot{\varphi}^2)^2}\biggr),
}
\end{equation}
{and the affine parameter can be eliminated by multiplying the right side of \eqref{angular_momentum} with the left side of~\eqref{energy_identity}. This leads to}
\begin{equation}\label{angular_momentum_equatorial_plane_without_affine_parameter}
{
L
=-\dfrac{E}{h_{tt}}\dfrac{1}{1+\phi_0}
r^2\dfrac{d\varphi}{dt} \biggl(1+\phi_2\dfrac{h_{rr}^2\biggl(\dfrac{dr}{dt}\biggr)^4}{\biggl(h_{rr}\biggl(\dfrac{dr}{dt}\biggr)^2+r^2\biggl(\dfrac{d\varphi}{dt}\biggr)^2\biggr)^2}\biggr).
}
\end{equation}

{Linearising \eqref{angular_momentum_equatorial_plane_without_affine_parameter} with respect to the perturbation functions (especially, replacing $(1+\phi_0)^{-1}$ with its linear approximation $1 - \phi_0$) then results in}
\begin{equation}\label{conserved_quantity_2}
L = -\dfrac{{E}}{h_{tt}} {r^2}\dfrac{d\varphi}{dt}\biggl(1 - \phi_0 + \phi_2\dfrac{h_{rr}^2\biggl(\dfrac{dr}{dt}\biggr)^{{4}}}{{\biggl(}h_{rr}\biggl(\dfrac{dr}{dt}\biggr)^2 + r^2\biggl(\dfrac{d\varphi}{dt}\biggr)^2{\biggr)^2}}\biggr).
\end{equation}

Since there are no derivatives with respect to the affine parameter in \eqref{conserved_quantity_1} and \eqref{conserved_quantity_2}, from now on we will designate derivatives with respect to $t$ by an overdot. Derivatives with respect to $r$ will be designated by a prime. Furthermore, with this we define the ``speed''
\begin{equation}
u := \sqrt{h_{rr}\dot{r}^2 + r^2\dot{\varphi}^2},
\end{equation}
the angle $\alpha$ uniquely determined by
\begin{equation}\label{alpha_definition}
\cos\alpha = \sqrt{h_{rr}}\dfrac{\dot{r}}{u}\text{ and }\sin\alpha = r\dfrac{\dot{\varphi}}{u}
\end{equation}
and the coefficients
\begin{equation}
p^{ik} := \cos^i\alpha\sin^k\alpha\ (i,\ k\ \in \mathbb{N}_0).
\end{equation}

Equations \eqref{conserved_quantity_1} and \eqref{conserved_quantity_2} then read
\begin{equation}\label{def_H}
2\dfrac{\mathcal{L}}{E^2} = \dfrac{{1}}{h_{tt}} + \dfrac{u^2}{h_{tt}^2}
- \phi_0\biggl(\dfrac{{1}}{h_{tt}}+2\dfrac{u^2}{h_{tt}^2}\biggr)
+ \phi_1\dfrac{h_{rr}}{h_{tt}^2}\dot{r}^2
+ \phi_2\dfrac{u^2}{h_{tt}^2} p^{22}=:H(r,\ \dot{r},\ \dot{\varphi})
\end{equation}
and
\begin{equation}\label{def_Lambda}
-\dfrac{L}{E} = \dfrac{r^2}{h_{tt}}\dot{\varphi}\biggl(1 - \phi_0 + \phi_2p^{40}\biggr)=:\Lambda(r,\ \dot{r},\ \dot{\varphi}).
\end{equation}

By differentiation of \eqref{def_H} and \eqref{def_Lambda} with respect to $t$ we get rid of the constants of motion and obtain the linear (in the algebraic sense) system of equations
\begin{equation}
\begin{split}
\dfrac{\partial H}{\partial\dot{r}}\ddot{r} + \dfrac{\partial H}{\partial\dot{\varphi}}\ddot{\varphi}
= &-\dfrac{\partial H}{\partial r}\dot{r},\\
\dfrac{\partial\Lambda}{\partial\dot{r}}\ddot{r} + \dfrac{\partial\Lambda}{\partial\dot{\varphi}}\ddot{\varphi}
= &-\dfrac{\partial\Lambda}{\partial r}\dot{r},
\end{split}
\end{equation}
that is to be solved for $\ddot{r}$ and $\ddot{\varphi}$.

After a very involved but elementary calculation we arrive at the following equations of motion, again linearized with respect to the $\phi_n$:
\begin{equation}
\begin{split}
\ddot{r} = &\dfrac{-c^2}{2}\biggl(\dfrac{r_S}{r^2}-\dfrac{r_S^2}{r^3}\biggr) 
+ \dfrac{3}{2r}\dfrac{r_S}{r-r_S}\dot{r}^2
+ (r-r_S)\dot{\varphi}^2
- \dfrac{c^2}{2}\biggl(\dfrac{r_S}{r^2}-\dfrac{r_S^2}{r^3}\biggr)\phi_0\\
&+ \biggr(\dot{r}^2-\dfrac{c^2}{2}\biggr(1-\dfrac{r_S}{r}\biggr)^2\biggr)\phi_0^\prime
+ \biggr(\dfrac{c^2}{2}\biggr(\dfrac{r_S}{r^2} - \dfrac{r_S^2}{r^3}\biggr) - (r-r_S)\dot{\varphi}^2\biggr)\phi_1
- \dfrac{\dot{r}^2}{2}\phi_1^\prime\\
&+\biggl\{-\dfrac{c^2}{2}\biggl(\dfrac{r_S}{r^2}-\dfrac{r_S^2}{r^3}\biggr) \biggl(4p^{24}-p^{04}\biggr)
+ \dot{u}^2\biggl(\biggl(\dfrac{1}{r}-\dfrac{r_S}{r^2}\biggr)\biggl(2p^{22}-p^{02}\biggr)
+ \biggl(\dfrac{4}{r}-\dfrac{9}{2}\dfrac{r_S}{r^2}\biggr)p^{24}\\
&+ \dfrac{r_S}{2r^2}p^{24}\biggr)\biggr\}\phi_2
+ \dfrac{1}{2}\biggl(1-\dfrac{r_S}{r}\biggr)u^2\biggl(p^{42}-p^{24}\biggr)\phi_2^\prime,
\end{split}
\end{equation}
\begin{equation}\label{ddot_varphi}
\ddot{\varphi} = \dfrac{3r_S-2r}{r-r_S}\dfrac{\dot{r}\dot{\varphi}}{r}
+ \dot{r}\dot{\varphi}\phi_0^\prime
- \biggl(\dfrac{4}{r}-2\dfrac{r_S}{r^2}\dfrac{c^2}{u^2}\biggr)\dot{r}\dot{\varphi}p^{22}\phi_2-\dot{r}\dot{\varphi}p^{40}\phi_2^\prime.
\end{equation}

\section{Equations for the Three-Dimensional Simulations}\label{app:3d_equations}

\subsection{Lagrangian in Isotropic Coordinates}

For the numerical study of observable effects of Finslerian perturbations on earth-bound satellites in the general geometry, without a common orbital plane of the satellite and the ground station, we used isotropic coordinates $t$, $x$, $y$, and $z$. These coordinates treat all directions in space and all orbital planes on an equal footing and are routinely used in orbit calculation programs. Schwarzschild coordinates and isotropic coordinates are related by the equations
\begin{equation}\label{31d}
x=\rho\sin\vartheta\cos\varphi,\ y=\rho\sin\vartheta\sin\varphi,\ z=\rho\cos\vartheta,\ r=\rho\biggl(1+\frac{r_S}{4\rho}\biggr)^2.
\end{equation}

In the following a prime denotes a derivative with respect to $\rho$ (e.g., $r^\prime:=dr/d\rho$), an overdot denotes a derivative with respect to the affine parameter $s$ (e.g., $\dot{r}:=dr/ds$), and a dot between two three-vectors denotes their Euclidean inner product (e.g., $\mathbf{r}\cdot\dot{\mathbf{r}}:=x\dot{x}+y\dot{y}+z\dot{z}$). Latin indices take on values 1, 2 or 3. Greek indices take on values 0, 1, 2 or 3. From \eqref{31d} and with the definitions
\begin{equation}
x^0:=t,\ x^1:=x,\ x^2:=y,\ x^3:=z,\ \mathbf{r}:=(x^k)
\end{equation}
it follows
\begin{equation}
\rho=\sqrt{\mathbf{r}\cdot\mathbf{r}},\ \dot{\rho}=\frac{\mathbf{r}\cdot\dot{\mathbf{r}}}{\rho},
\end{equation}
\begin{equation}
r^\prime=\biggl(1+\frac{r_S}{4\rho}\biggr)\biggl(1-\frac{r_S}{4\rho}\biggr),
\end{equation}
\begin{equation}
\dot{r}=r^\prime\dot{\rho}=\biggl(1+\frac{r_S}{4\rho}\biggr)\biggl(1-\frac{r_S}{4\rho}\biggr)\frac{\mathbf{r}\cdot\dot{\mathbf{r}}}{\rho}.
\end{equation}

From the metric coefficients of the Schwarzschild coordinates and of the isotropic coordinates
\begin{equation}
h_{rr}:=\biggl(1-\frac{r_S}{r}\biggr)^{-1}=\biggl(1+\frac{r_S}{4\rho}\biggr)^2\biggl(1-\frac{r_S}{4\rho}\biggr)^{-2},
\end{equation}
\begin{equation}
h_{tt}:=\frac{-c^2}{h_{rr}},
\end{equation}
\begin{equation}
h_{\rho\rho}:=\biggl(1+\frac{r_S}{4\rho}\biggr)^4
\end{equation}
and \eqref{31d} it follows
\begin{equation}
h_{rr}\dot{r}^2=h_{\rho\rho}\dot{\rho}^2,
\end{equation}
\begin{equation}
r^2=h_{\rho\rho}\rho^2.
\end{equation}

If we restrict the analysis for a moment to geodesics in the plane $\vartheta=\pi/2$, the following equations for $\dot{\varphi}$ can be derived from \eqref{31d}:
\begin{equation}
\dot{\varphi}=\frac{x\dot{y}-y\dot{x}}{x^2+y^2},\ \dot{\varphi}^2=\frac{(x^2+y^2)(\dot{x}^2+\dot{y}^2)-(x\dot{x}+y\dot{y})^2}{(x^2+y^2)^2}=\frac{\dot{\mathbf{r}}\cdot\dot{\mathbf{r}}}{\rho^2}-\frac{({\mathbf{r}}\cdot\dot{\mathbf{r}})^2}{\rho^4}.
\end{equation}

With the above results we are now prepared to transform the Lagrangian \eqref{Lagrangian_A}, restricted to the plane $\vartheta=\pi/2$, to isotropic coordinates:
\begin{equation}\label{31p}
\begin{split}
2\mathcal{L}
=&\ (1+\phi_0)h_{tt}\dot{t}^2
+(1+\phi_1)h_{\rho\rho}\dot{\rho}^2
+h_{\rho\rho}\rho^2\dot{\varphi}^2
+\frac{\phi_2\rho^2h_{\rho\rho}\dot{\rho}^2\dot{\varphi}^2}{\dot{\rho}^2+\rho^2\dot\varphi^2}\\
=&\ (1+\phi_0)h_{tt}\dot{t}^2
+h_{\rho\rho}\biggl(
\rho^2\dot{\varphi}^2
+(1+\phi_1)\dot{\rho}^2
+\frac{\phi_2\rho^2\dot{\rho}^2\dot{\varphi}^2}{\dot{\rho}^2+\rho^2\dot\varphi^2}\biggr)\\
=&\ (1+\phi_0)h_{tt}\dot{t}^2
+h_{\rho\rho}\biggl(
\dot{\mathbf{r}}\cdot\dot{\mathbf{r}}-\frac{({\mathbf{r}}\cdot\dot{\mathbf{r}})^2}{\rho^2}
+(1+\phi_1)\frac{(\mathbf{r}\cdot\dot{\mathbf{r}})^2}{\rho^2}
+\frac{\phi_2\rho^2\dot{\rho}^2\dot{\varphi}^2}{\dot{\rho}^2+\rho^2\dot\varphi^2}\biggr)\\
=&\ (1+\phi_0)h_{tt}\dot{t}^2
+h_{\rho\rho}\biggl(
\dot{\mathbf{r}}\cdot\dot{\mathbf{r}}
+\frac{(\mathbf{r}\cdot\dot{\mathbf{r}})^2}{\rho^2}\phi_1
+\frac{\rho^2\dot{\rho}^2\dot{\varphi}^2}{\dot{\rho}^2+\rho^2\dot\varphi^2}\phi_2\biggr)\\
=&\ (1+\phi_0)h_{tt}\dot{t}^2
+h_{\rho\rho}\biggl(
\dot{\mathbf{r}}\cdot\dot{\mathbf{r}}
+\frac{(\mathbf{r}\cdot\dot{\mathbf{r}})^2}{\rho^2}\phi_1
+\frac{\rho^2\dot{\rho}^2\dot{\varphi}^2}{\dot{\mathbf{r}}\cdot\dot{\mathbf{r}}}\phi_2\biggr)\\
=&\ (1+\phi_0)h_{tt}\dot{t}^2
+h_{\rho\rho}\biggl(
\dot{\mathbf{r}}\cdot\dot{\mathbf{r}}
+\frac{(\mathbf{r}\cdot\dot{\mathbf{r}})^2}{\rho^2}\phi_1
+\frac{\rho^2}{\dot{\mathbf{r}}\cdot\dot{\mathbf{r}}} \frac{(\mathbf{r}\cdot\dot{\mathbf{r}})^2}{\rho^2} \biggl(\frac{\dot{\mathbf{r}}\cdot\dot{\mathbf{r}}}{\rho^2}-\frac{({\mathbf{r}}\cdot\dot{\mathbf{r}})^2}{\rho^4}\biggr) \phi_2\biggr)\\
=&\ (1+\phi_0)h_{tt}\dot{t}^2
+h_{\rho\rho}\biggl(
\dot{\mathbf{r}}\cdot\dot{\mathbf{r}}
+\frac{(\mathbf{r}\cdot\dot{\mathbf{r}})^2}{\rho^2}(\phi_1+\phi_2)
-\frac{(\mathbf{r}\cdot\dot{\mathbf{r}})^4}{\rho^4\dot{\mathbf{r}}\cdot\dot{\mathbf{r}}}\phi_2\biggr)\\
=&\ (1+\phi_0)h_{tt}\dot{t}^2
+h_{\rho\rho}\dot{\mathbf{r}}\cdot\dot{\mathbf{r}}\biggl(
1
+\frac{(\mathbf{r}\cdot\dot{\mathbf{r}})^2}{\rho^2\dot{\mathbf{r}}\cdot\dot{\mathbf{r}}}(\phi_1+\phi_2)
-\frac{(\mathbf{r}\cdot\dot{\mathbf{r}})^4}{\rho^4(\dot{\mathbf{r}}\cdot\dot{\mathbf{r}})^2}\phi_2\biggr).
\end{split}
\end{equation}

By use of the definitions
\begin{equation}
f:=\frac{(\mathbf{r}\cdot\dot{\mathbf{r}})^2}{(\mathbf{r}\cdot\mathbf{r})(\dot{\mathbf{r}}\cdot\dot{\mathbf{r}})},
\end{equation}
\begin{equation}
\phi_3:=\phi_1+\phi_2
\end{equation}
the last line of \eqref{31p} can be written in the more compact form
\begin{equation}\label{Lagrangian_1}
2\mathcal{L}=h_{tt}(1+\phi_0)\dot{t}^2
+h_{\rho\rho}(1+f\phi_3-f^2\phi_2)\dot{\mathbf{r}}\cdot\dot{\mathbf{r}}.
\end{equation}

By use of Lagrange's identity $(\mathbf{a}\times\mathbf{b})\cdot(\mathbf{c}\times\mathbf{d})=(\mathbf{a}\cdot\mathbf{c})(\mathbf{b}\cdot\mathbf{d})-(\mathbf{b}\cdot\mathbf{c})(\mathbf{a}\cdot\mathbf{d})$ the last line of \eqref{31p} can also be written as
\begin{equation}\label{Lagrangian_2}
2\mathcal{L}
=h_{tt}\dot{t}^2(1+\phi_0)+h_{\rho\rho}\dot{\mathbf{r}}\cdot\dot{\mathbf{r}}
\biggl(
1+f\phi_1+
f\frac{(\mathbf{r}\times\dot{\mathbf{r}})\cdot(\mathbf{r}\times\dot{\mathbf{r}})}{(\mathbf{r}\cdot\mathbf{r})(\dot{\mathbf{r}}\cdot\dot{\mathbf{r}})}\phi_2
\biggr).
\end{equation}

Equations \eqref{Lagrangian_1} and \eqref{Lagrangian_2} have been derived for geodesics in the plane $\vartheta=\pi/2$, {but due to their spherical symmetry} they are invariant under spatial rotations of the isotropic coordinate system and therefore valid for arbitrary geodesics.

{To determine the equation for the coordinate light speed $\overline{c}(\mathbf{r},\ \mathbf{d})$ at the position $\mathbf{r}$ in the direction $\mathbf{d}$, we first note for the Lagrangian \eqref{Lagrangian_1} and geodesics with $\dot{t}\neq0$ (e.g., causal geodesics) the identity}
\begin{equation}\label{geodesic_coordinate_velocity}
{\mathcal{L}(t,\ \mathbf{r},\ \dot{t},\ \dot{\mathbf{r}}) = \dot{t}^2 \mathcal{L}(t,\ \mathbf{r},\ 1,\ \frac{d\mathbf{r}}{dt}).}
\end{equation}

{A null geodesic crossing the position $\mathbf{r}=\lvert\mathbf{r}\rvert\mathbf{\hat{r}}$ in the direction $\mathbf{d}=\lvert\mathbf{d}\rvert\mathbf{\hat{d}}$ thus fulfils the equation}
\begin{equation}\label{lightspeed2}
{0 = \mathcal{L}(t,\ \mathbf{r},\ 1,\ \overline{c}(\mathbf{r},\ \mathbf{d}) \mathbf{\hat{d}}).}
\end{equation}

{Inserting} the Lagrangian \eqref{Lagrangian_1} {into \eqref{lightspeed2}}  and solving the resulting equation for the coordinate light speed $\overline{c}${$(\mathbf{r},\ \mathbf{d})$} {then} gives
\begin{equation}\label{lightspeed}
\overline{c}(\mathbf{r},\ \mathbf{d})=\sqrt{\frac{-h_{tt}}{h_{\rho\rho}}\frac{1+\phi_0}{1+\phi_3(\mathbf{\hat{r}}\cdot\mathbf{\hat{d}})^2-\phi_2(\mathbf{\hat{r}}\cdot\mathbf{\hat{d}})^4}}.
\end{equation}

{Using \eqref{geodesic_coordinate_velocity} for a timelike geodesic with proper time $\tau$ as the affine parameter results in}
\begin{equation}\label{geodesic_tau}
{-\frac{c^2}{2} = \mathcal{L}(t,\ \mathbf{r},\ \frac{dt}{d\tau},\ \frac{d\mathbf{r}}{d\tau}) = \biggl(\frac{dt}{d\tau}\biggr)^2 \mathcal{L}(t,\ \mathbf{r},\ 1,\ \frac{d\mathbf{r}}{dt}).}
\end{equation}

{Inserting} the Lagrangian \eqref{Lagrangian_1} {into \eqref{geodesic_tau}} {results in} the following equation for the proper times of the ground station and the satellite:
\begin{equation}\label{proper_time}
\frac{d\tau}{dt}=\frac{1}{c}\sqrt{-h_{tt}(1+\phi_0)-h_{\rho\rho}v^2(1+f\phi_3-f^2\phi_2)}.
\end{equation}

\subsection{Equations of Motion in Isotropic Coordinates}

The further analysis is based on the Lagrangian \eqref{Lagrangian_1}. As we are only interested in causal geodesics, we will only consider geodesics with $\dot{t}\neq0$. We continue with several definitions{.}

{For scalar functions $b(...,\ \mathbf{s},\ ...)$ and vector functions $\mathbf{b}(...,\ \mathbf{s},\ ...)$ with a vector argument $\mathbf{s}$ we will use the notations}
\begin{equation}
{\frac{\partial b}{\partial\mathbf{s}} := \biggl(\frac{\partial b}{\partial s^l}\biggr),\ \frac{\partial\mathbf{b}}{\partial\mathbf{s}} := \biggl(\frac{\partial b^k}{\partial s^l}\biggr).}
\end{equation}

{Furthermore, we define}
\begin{equation}
v^k:=\frac{dx^k}{dt},\ \mathbf{v}:=(v^k),\ a^k:=\frac{dv^k}{dt},\ \mathbf{a}:=(a^k),
\end{equation}
\begin{equation}
v:=\sqrt{\mathbf{v}\cdot\mathbf{v}},\ \xi:=\mathbf{r}\cdot\mathbf{v},
\end{equation}
and note that the affine parameter $s$ can be eliminated from the function $f$ in the following way: 
\begin{equation}\label{f_of_t}
f=\frac{\xi^2}{\rho^2v^2}.
\end{equation}

From the Euler-Lagrange equation
\begin{equation}
\frac{d}{ds}\frac{\partial\mathcal{L}}{\partial\dot{t}}=\frac{\partial\mathcal{L}}{\partial t}
\end{equation}
and
\begin{equation}
\frac{\partial\mathcal{L}}{\partial t}=0,
\end{equation}
\begin{equation}\label{energy_1}
-E:=\frac{\partial\mathcal{L}}{\partial\dot{t}}=h_{tt}(1+\phi_0)\dot{t},
\end{equation}
we get
\begin{equation}
\begin{split}
0 = -\dot{E} =&\ \dot{h_{tt}}(1+\phi_0)\dot{t} + h_{tt}\dot{\phi_0}\dot{t} + h_{tt}(1+\phi_0)\ddot{t}\\
=&\ h_{tt}^\prime\frac{d\rho}{dt}(1+\phi_0)\dot{t}^2 + h_{tt}\phi_0^\prime\frac{d\rho}{dt}\dot{t}^2 + h_{tt}(1+\phi_0)\ddot{t}
\end{split}
\end{equation}
{and therefore}
\begin{equation}\label{03a}
w:=\frac{\ddot{t}}{\dot{t}^2}{=-\biggl(\frac{h_{tt}^\prime}{h_{tt}}+\frac{\phi_0^\prime}{1+\phi_0}\biggr)\frac{d\rho}{dt}}=-\biggl(\frac{h_{tt}^\prime}{h_{tt}}+\frac{\phi_0^\prime}{1+\phi_0}\biggr)\frac{\xi}{\rho}.
\end{equation}

By use of
\begin{equation}
\frac{\partial f}{\partial\mathbf{r}}
=2f\biggl(\frac{\mathbf{v}}{\xi}-\frac{\mathbf{r}}{\rho^2}\biggr)
\end{equation}
we get
\begin{equation}\label{23b}
\begin{split}
\mathbf{u}:=&\ \frac{\partial\mathcal{L}}{\partial\mathbf{r}}\dot{t}^{-2}\\
=&\ \frac{\mathbf{r}}{2\rho}(h_{tt}^\prime(1+\phi_0)+h_{tt}\phi_0^\prime)
+\frac{\mathbf{r}}{2\rho}h_{\rho\rho}^\prime v^2(1+f\phi_3-f^2\phi_2)\\
&\ +\frac{\mathbf{r}}{2\rho}h_{\rho\rho}v^2(f\phi_3^\prime-f^2\phi_2^\prime)
+\frac{1}{2}h_{\rho\rho}v^2\frac{\partial f}{\partial\mathbf{r}}(\phi_3-2f\phi_2)\\
=&\ \frac{\mathbf{r}}{2\rho}(h_{tt}^\prime(1+\phi_0)+h_{tt}\phi_0^\prime)
+\frac{\mathbf{r}}{2\rho}h_{\rho\rho}^\prime v^2(1+f\phi_3-f^2\phi_2)\\
&\ +\frac{\mathbf{r}}{2\rho}h_{\rho\rho}v^2(f\phi_3^\prime-f^2\phi_2^\prime)
+h_{\rho\rho}\frac{\xi}{\rho^2}\biggl(\mathbf{v}-\frac{\xi}{\rho^2}\mathbf{r}\biggr)(\phi_3-2f\phi_2)\\
=&\ (h_{tt}^\prime(1+\phi_0)+h_{tt}\phi_0^\prime
+h_{\rho\rho}^\prime v^2(1+f\phi_3-f^2\phi_2)
+h_{\rho\rho}v^2(f\phi_3^\prime-f^2\phi_2^\prime))\frac{\mathbf{r}}{2\rho}\\
&\ +h_{\rho\rho}(\phi_3-2f\phi_2)\frac{\xi}{\rho^2}\biggl(\mathbf{v}-\frac{\xi}{\rho^2}\mathbf{r}\biggr).
\end{split}
\end{equation}

By use of
\begin{equation}
\frac{\partial f}{\partial\mathbf{\dot{r}}}
=2f\biggl(\frac{\mathbf{r}}{\xi}-\frac{\mathbf{v}}{v^2}\biggr)\dot{t}^{-1}
\end{equation}
we get
\vspace{12pt}
\begin{equation}\label{23a}
\begin{split}
\mathbf{w}:=&\ \frac{\partial\mathcal{L}}{\partial\dot{\mathbf{r}}}\dot{t}^{-1}\\
=&\ h_{\rho\rho}\dot{\mathbf{r}}(1+f\phi_3-f^2\phi_2)\dot{t}^{-1}
+\frac{1}{2}h_{\rho\rho}\dot{\mathbf{r}}\cdot\dot{\mathbf{r}}(\phi_3-2f\phi_2)\frac{\partial f}{\partial\dot{\mathbf{r}}}\dot{t}^{-1}\\
=&\ h_{\rho\rho}\dot{\mathbf{r}}(1+f\phi_3-f^2\phi_2)\dot{t}^{-1}
+h_{\rho\rho}\dot{\mathbf{r}}\cdot\dot{\mathbf{r}}(\phi_3-2f\phi_2)
f\biggl(\frac{\mathbf{r}}{\xi}-\frac{\mathbf{v}}{v^2}\biggr)\dot{t}^{-2}\\
=&\ h_{\rho\rho}\mathbf{v}(1+f\phi_3-f^2\phi_2)
+h_{\rho\rho}(\phi_3-2f\phi_2)\biggl(\frac{\xi}{\rho^2}\mathbf{r}-f\mathbf{v}\biggr)\\
=&\ h_{\rho\rho}\mathbf{v}(1+f\phi_3-f^2\phi_2)
+h_{\rho\rho}(\phi_3-2f\phi_2)\frac{\xi}{\rho^2}\mathbf{r}
-h_{\rho\rho}(\phi_3-2f\phi_2)f\mathbf{v}\\
=&\ h_{\rho\rho}\mathbf{v}(1+f\phi_3-f^2\phi_2)
-h_{\rho\rho}(\phi_3-2f\phi_2)f\mathbf{v}
+h_{\rho\rho}(\phi_3-2f\phi_2)\frac{\xi}{\rho^2}\mathbf{r}\\
=&\ h_{\rho\rho}(1+f^2\phi_2)\mathbf{v}+h_{\rho\rho}(\phi_3-2f\phi_2)\frac{\xi}{\rho^2}\mathbf{r}\\
=&\ h_{\rho\rho}\biggl((\phi_3-2f\phi_2)\frac{\xi}{\rho^2}\mathbf{r}+(1+f^2\phi_2)\mathbf{v}\biggr).
\end{split}
\end{equation}

In the following we will make use of the definitions
\begin{equation}\label{23z}
M^{kl}:=\frac{\partial w^k}{\partial v^l},\ M:=(M^{kl}),\ N^{kl}:=\frac{\partial w^k}{\partial x^l},\ N:=(N^{kl}).
\end{equation}

From the Euler-Lagrange equations and \eqref{03a}, \eqref{23b}, \eqref{23a} and \eqref{23z} we get
\begin{equation}\label{08y}
\begin{split}
0=\ &\dot{t}^{-2}\biggl(\frac{d}{ds}\frac{\partial\mathcal{L}}{\partial\dot{\mathbf{r}}}
-\frac{\partial\mathcal{L}}{\partial\mathbf{r}}\biggr)
=\dot{t}^{-2}\biggl(\frac{d}{ds}(\dot{t}\mathbf{w})
-\frac{\partial\mathcal{L}}{\partial\mathbf{r}}\biggr)
=\frac{\ddot{t}}{\dot{t}^2}\mathbf{w}+\frac{d\mathbf{w}}{dt}
-\dot{t}^{-2}\frac{\partial\mathcal{L}}{\partial\mathbf{r}}\\
=\ &\frac{\ddot{t}}{\dot{t}^2}\mathbf{w}
+\frac{\partial\mathbf{w}}{\partial\mathbf{r}}\mathbf{v}
+\frac{\partial\mathbf{w}}{\partial\mathbf{v}}\mathbf{a}
-\dot{t}^{-2}\frac{\partial\mathcal{L}}{\partial\mathbf{r}}
=w\mathbf{w}
+N\mathbf{v}
+M\mathbf{a}
-\mathbf{u}.
\end{split}
\end{equation}

From \eqref{08y} the acceleration vector $\mathbf{a}$ can be calculated as
\begin{equation}\label{23x}
\mathbf{a}=M^{-1}(\mathbf{u}-w\mathbf{w}-N\mathbf{v}).
\end{equation}

In the next step we need to calculate $M$ and $N$:
\begin{equation}
M^{kl}=h_{\rho\rho}\biggl((1+f^2\phi_2)\delta^{kl}+(\phi_3-6f\phi_2)\frac{x^k x^l}{\rho^2}+4f^2\phi_2\biggl(\frac{x^k v^l+v^k x^l}{\xi}-\frac{v^k v^l}{v^2}\biggr)\biggr),
\end{equation}
\begin{equation}
\begin{split}
N^{kl}=&\frac{h_{\rho\rho}}{\rho^2}\biggl((\phi_3-2f\phi_2)\xi\delta^{kl}
+\biggl(8f\frac{\phi_2}{\rho}-2\frac{\phi_3}{\rho}+\phi_3^\prime-2f\phi_2^\prime\biggr)\xi\frac{x^k x^l}{\rho}
+(\phi_3-6f\phi_2)x^k v^l\biggr)\\
&+h_{\rho\rho}f^2\biggl(
\biggl(\phi_2^\prime-4\frac{\phi_2}{\rho}\biggr)\frac{v^k x^l}{\rho}
+4\phi_2\frac{v^k v^l}{\xi}\biggr)
+\frac{h_{\rho\rho}^\prime}{\rho}\biggl((\phi_3-2f\phi_2)\xi\frac{x^k x^l}{\rho^2}+(1+f^2\phi_2)v^k x^l\biggr).
\end{split}
\end{equation}

As we assume the Finslerian perturbations to be small in the physical situations that are considered in this study, it suffices to calculate the right side of \eqref{23x} only to the first order of the $\phi_k$ and their derivatives. The result is
\begin{equation}
(M^{-1})^{kl} = \frac{1}{h_{\rho\rho}}\biggl((1-f^2\phi_2)\delta^{kl}+(6f\phi_2-\phi_3)\frac{x^k x^l}{\rho^2}+
4f^2\phi_2\biggl(\frac{v^k v^l}{v^2}-\frac{x^k v^l+v^k x^l}{\xi}\biggr)\biggr),
\end{equation}
\begin{equation}\label{11b}
\begin{split}
M^{-1}\mathbf{u} = &
\biggl\lbrace\frac{h_{\rho\rho}^\prime}{h_{\rho\rho}}v^2(1+(f-1)\phi_3+6(f-f^2)\phi_2)+\frac{h_{tt}^\prime}{h_{\rho\rho}}(1+\phi_0-\phi_3+(6f-5f^2)\phi_2)\\
&+\frac{h_{tt}}{h_{\rho\rho}}\phi_0^\prime+2\frac{\xi^2}{\rho^3}(2f\phi_2-\phi_3)+v^2f(\phi_3^\prime-f\phi_2^\prime)\biggr\rbrace\frac{\mathbf{r}}{2\rho}\\
&+\phi_3\frac{\xi}{\rho^2}\mathbf{v}+2f\biggl\lbrace\frac{f-1}{h_{\rho\rho}}\biggl(\frac{h_{tt}^\prime}{v^2}+h_{\rho\rho}^\prime\biggr)-\frac{1}{\rho}\biggr\rbrace\phi_2\frac{\xi}{\rho}\mathbf{v},
\end{split}
\end{equation}
\begin{equation}\label{11c}
\begin{split}
M^{-1}N\mathbf{v} = &
\biggl\lbrace(1-2f)\frac{\phi_3}{\rho}+(8f^2-6f)\frac{\phi_2}{\rho}+f\phi_3^\prime-2f^2\phi_2^\prime\biggr\rbrace\frac{v^2}{\rho}\mathbf{r}\\
&+\biggl\lbrace\frac{h_{\rho\rho}^\prime}{h_{\rho\rho}}+\frac{\phi_3}{\rho}+(2f-4f^2)\frac{\phi_2}{\rho}+f^2\phi_2^\prime\biggr\rbrace\frac{\xi}{\rho}\mathbf{v},
\end{split}
\end{equation}
\begin{equation}\label{11d}
M^{-1}w\mathbf{w} = -\biggl(\frac{h_{tt}^\prime}{h_{tt}}+\phi_0^\prime\biggr)\frac{\xi}{\rho}\mathbf{v}.
\end{equation}

From \eqref{23x}, \eqref{11b}, \eqref{11c}, and \eqref{11d} the acceleration vector $\mathbf{a}$ is calculated as
\begin{equation}\label{acceleration_vector_a}
\begin{split}
\mathbf{a} = &\biggl\lbrace\frac{h_{\rho\rho}^\prime}{h_{\rho\rho}}v^2+\frac{h_{tt}^\prime}{h_{\rho\rho}}(1+\phi_0-\phi_3+(6f-5f^2)\phi_2)\\
&+\biggl(\frac{h_{\rho\rho}^\prime}{h_{\rho\rho}}+\frac{2}{\rho}\biggr)v^2(f-1)(\phi_3-6f\phi_2)
 +\frac{h_{tt}}{h_{\rho\rho}}\phi_0^\prime+v^2(3f^2\phi_2^\prime-f\phi_3^\prime)\biggr\rbrace\frac{\mathbf{r}}{2\rho}\\
&+\biggl\lbrace\frac{h_{tt}^\prime}{h_{tt}}-\frac{h_{\rho\rho}^\prime}{h_{\rho\rho}}+2\biggl(\frac{h_{tt}^\prime}{v^2h_{\rho\rho}}+\frac{h_{\rho\rho}^\prime}{h_{\rho\rho}}+\frac{2}{\rho}\biggr)(f^2-f)\phi_2+\phi_0^\prime-f^2\phi_2^\prime\biggr\rbrace\frac{\xi}{\rho}\mathbf{v}.
\end{split}
\end{equation}

\subsection{Frequency Shift of the Signals}

We consider a general Finsler spacetime with a Lagrangian $\mathcal{L}(x,\ \dot{x})$ that is positively homogeneous of degree two with respect to $\dot{x}$ for all $\dot{x}\neq 0$, i.e.,
\begin{equation}
\mathcal{L}(x,\ k\dot{x})=k^2 \mathcal{L}(x,\ \dot{x})\hspace{0.6cm}\text{for all }k>0\text{ and all }\dot{x}\neq 0.
\end{equation}

This homogeneity condition of the Lagrangian implies that the Finsler metric
\begin{equation}
g_{\mu\nu}(x,\ \dot{x}):=\frac{\partial^2\mathcal{L}(x,\ \dot{x})}{\partial\dot{x}^\mu\partial\dot{x}^\nu}
\end{equation}
is positively homogeneous of degree zero with respect to $\dot{x}$ for all $\dot{x}\neq 0$, i.e.,
\begin{equation}\label{Finsler_metric_homogeneity condition}
g_{\mu\nu}(x,\ k\dot{x})=g_{\mu\nu}(x,\ \dot{x})\hspace{0.6cm}\text{for all }k>0\text{ and all }\dot{x}\neq 0.
\end{equation}

In this Finsler spacetime we consider an emitter with worldline $\alpha$ and proper time $\tau_\alpha$, a receiver with worldline $\sigma$ and proper time $\tau_\sigma$, and a light ray $\lambda$ (which describes the signal from the satellite to the ground station) with affine parameter $s$, connecting $\alpha$ and $\sigma$. The light ray is emitted at event 1 and received at event 2. Then, according to~\cite{reference:Hasse_2019}, Equation (33), the relative frequency shift of the light ray is
\begin{equation}\label{frequency_shift_1}
 \dfrac{f_2}{f_1}
=\dfrac{g_{\kappa\nu}(\lambda(s_2),\ \dot{\lambda}(s_2)) \dot{\lambda}^\nu(s_2) \dot{\sigma}^\kappa(\tau_{\sigma_2})}
       {g_{\kappa\nu}(\lambda(s_1),\ \dot{\lambda}(s_1)) \dot{\lambda}^\nu(s_1) \dot{\alpha}^\kappa(\tau_{\alpha_1})}.
\end{equation}

Now the general formula \eqref{frequency_shift_1} is applied to the Lagrangian \eqref{Lagrangian_1} and the derivative with respect to the affine parameter $s$ is replaced by the derivative with respect to the coordinate time $t$. We first note the equation
\begin{equation}\label{energy_2}
\dfrac{d\lambda^0}{ds}=\frac{-E_\lambda}{h_{tt}(1+\phi_0)}
\end{equation}
that follows from \eqref{energy_1}. $E_\lambda$ denotes the constant ``energy'' of the light ray. To make use of the homogeneity property \eqref{Finsler_metric_homogeneity condition} of the Finsler metric, the light ray is assumed to fulfil the condition
\begin{equation}\label{dtds_gt_0}
\dfrac{d\lambda^0}{ds}>0.
\end{equation}

Combining \eqref{Finsler_metric_homogeneity condition}, \eqref{frequency_shift_1}, \eqref{energy_2} and \eqref{dtds_gt_0} gives for the frequency shift the equation
\begin{equation}\label{frequency_shift_2}
\begin{split}
	\dfrac{f_2}{f_1}
=&\dfrac{g_{\kappa\nu}\biggl(\lambda(s_2),\ \dfrac{d\lambda^0}{ds}(s_2)\dfrac{d\lambda}{dt}(t_2)\biggr)
				\dfrac{d\lambda^0}{ds}(s_2)
				\dfrac{d\lambda^\nu}{dt}(t_2)
				\dfrac{d\sigma^0}{d\tau_\sigma}(\tau_{\sigma_2})
				\dfrac{d\sigma^\kappa}{dt}(t_2)}
				{g_{\kappa\nu}\biggl(\lambda(s_1),\ \dfrac{d\lambda^0}{ds}(s_1)\dfrac{d\lambda}{dt}(t_1)\biggr)
				\dfrac{d\lambda^0}{ds}(s_1)
				\dfrac{d\lambda^\nu}{dt}(t_1)
				\dfrac{d\alpha^0}{d\tau_\alpha}(\tau_{\alpha_1})
				\dfrac{d\alpha^\kappa}{dt}(t_1)}\\
=&\dfrac{g_{\kappa\nu}\biggl(\lambda(s_2),\ \dfrac{d\lambda}{dt}(t_2)\biggr)
				\dfrac{d\lambda^0}{ds}(s_2)
				\dfrac{d\lambda^\nu}{dt}(t_2)
				\dfrac{d\tau_\alpha}{dt}(t_1)
				\dfrac{d\sigma^\kappa}{dt}(t_2)}
				{g_{\kappa\nu}\biggl(\lambda(s_1),\ \dfrac{d\lambda}{dt}(t_1)\biggr)
				\dfrac{d\lambda^0}{ds}(s_1)
				\dfrac{d\lambda^\nu}{dt}(t_1)
				\dfrac{d\tau_\sigma}{dt}(t_2)
				\dfrac{d\alpha^\kappa}{dt}(t_1)}\\
=&\dfrac{g_{\kappa\nu}\biggl(\lambda(s_2),\ \dfrac{d\lambda}{dt}(t_2)\biggr)
				\dfrac{-E_\lambda}{h_{tt}(\rho_2)(1+\phi_0(\rho_2))}
				\dfrac{d\lambda^\nu}{dt}(t_2)
				\dfrac{d\tau_\alpha}{dt}(t_1)
				\dfrac{d\sigma^\kappa}{dt}(t_2)}
				{g_{\kappa\nu}\biggl(\lambda(s_1),\ \dfrac{d\lambda}{dt}(t_1)\biggr)
				\dfrac{-E_\lambda}{h_{tt}(\rho_1)(1+\phi_0(\rho_1))}
				\dfrac{d\lambda^\nu}{dt}(t_1)
				\dfrac{d\tau_\sigma}{dt}(t_2)
				\dfrac{d\alpha^\kappa}{dt}(t_1)}\\
=&\dfrac{1+\phi_0(\rho_1)}{1+\phi_0(\rho_2)}
	\dfrac{h_{tt}(\rho_1)}{h_{tt}(\rho_2)}
	\dfrac{\dfrac{d\tau_\alpha}{dt}(t_1)}{\dfrac{d\tau_\sigma}{dt}(t_2)}
	\dfrac{g_{\kappa\nu}\biggl(\lambda(t_2),\ \dfrac{d\lambda}{dt}(t_2)\biggr)
				\dfrac{d\lambda^\nu}{dt}(t_2)
				\dfrac{d\sigma^\kappa}{dt}(t_2)}
				{g_{\kappa\nu}\biggl(\lambda(t_1),\ \dfrac{d\lambda}{dt}(t_1)\biggr)
				\dfrac{d\lambda^\nu}{dt}(t_1)
				\dfrac{d\alpha^\kappa}{dt}(t_1)}.
\end{split}
\end{equation}




\end{document}